\documentclass[lettersize,journal]{IEEEtran}
\usepackage{amsmath,amsfonts,lipsum}
\usepackage{algorithm}
\usepackage{array}
\usepackage{algpseudocode}
\usepackage[caption=false,font=normalsize,labelfont=sf,textfont=sf]{subfig}
\usepackage{textcomp}
\usepackage{stfloats}
\usepackage{cuted}%%stripsep-3pt
\usepackage{url}
\usepackage{verbatim}
\usepackage{graphicx}
\usepackage{cite}
\usepackage{amsthm}
\newcounter{mytempeqncnt}

\usepackage{color}
\usepackage{multicol}
\usepackage{tabularx}  % 关键包：自动列宽调整
\usepackage{ragged2e}  % 更好的对齐控制
\usepackage{xurl}      % 允许文献引用自动换行
\usepackage{booktabs}  % 美化表格线

\makeatletter

\newcommand{\Rmnum}[1]{\expandafter\@slowromancap\romannumeral #1@}
\makeatother

\begin{document}
\title{Hierarchical Learning for IRS-Assisted MEC Systems with Rate-Splitting Multiple Access}

\author{
        Yinyu Wu,
        Xuhui Zhang,
        Yingchao Jiao,
        Jinke Ren,
        Yanyan Shen,
        Bo Yang,
        Shuqiang Wang,
        and Dusit Niyato

\thanks{
Y. Wu and Y. Jiao are with the Shenzhen Institute of Advanced Technology, Chinese Academy of Sciences, Guangdong 518055, China, and also with the University of Chinese Academy of Sciences, Beijing 100049, China (e-mail: yg.wu@siat.ac.cn; yc.jiao@siat.ac.cn).
}

\thanks{
X. Zhang and J. Ren are with the Shenzhen Future Network of Intelligence Institute, the School of Science and Engineering, and the Guangdong Provincial Key Laboratory of Future Networks of Intelligence, The Chinese University of Hong Kong, Shenzhen, Guangdong 518172, China (e-mail: xu.hui.zhang@foxmail.com; jinkeren@cuhk.edu.cn).
}

\thanks{
Y. Shen and S. Wang are with the Shenzhen Institute of Advanced Technology, Chinese Academy of Sciences, Guangdong 518055, China, and also with Shenzhen University of Advanced Technology, Guangdong 518055, China (e-mail: yy.shen@siat.ac.cn; sq.wang@siat.ac.cn).
}

\thanks{B. Yang is with the Department of Automation and the Key Laboratory of System Control and Information Processing, Ministry of Education, Shanghai Jiao Tong University, Shanghai 200240, China (e-mail: bo.yang@sjtu.edu.cn).}

\thanks{D. Niyato is with the College of Computing and Data Science, Nanyang Technological University, Singapore 639798 (e-mail: dniyato@ntu.edu.sg).}

}

\maketitle

%摘要
\begin{abstract}
Intelligent reflecting surface (IRS)-assisted mobile edge computing (MEC) systems have shown notable improvements in efficiency, such as reduced latency, higher data rates, and better energy efficiency. However, the resource competition among users will lead to uneven allocation, increased latency, and lower throughput. Fortunately, the rate-splitting multiple access (RSMA) technique has emerged as a promising solution for managing interference and optimizing resource allocation in MEC systems. This paper studies an IRS-assisted MEC system with RSMA, aiming to jointly optimize the passive beamforming of the IRS, the active beamforming of the base station, the task offloading allocation, the transmit power of users, the ratios of public and private information allocation, and the decoding order of the RSMA to minimize the average delay from a novel uplink transmission perspective. Since the formulated problem is non-convex and the optimization variables are highly coupled, we propose a hierarchical deep reinforcement learning-based algorithm to optimize both continuous and discrete variables of the problem. Additionally, to better extract channel features, we design a novel network architecture within the policy and evaluation networks of the proposed algorithm, combining convolutional neural networks and densely connected convolutional network for feature extraction. Simulation results indicate that the proposed algorithm not only exhibits excellent convergence performance but also outperforms various benchmarks.
\end{abstract}

%索引关键词
\begin{IEEEkeywords}
Intelligent reflecting surface, rate-splitting multiple access, mobile edge computing, deep reinforcement learning.
\end{IEEEkeywords}

%介绍
\section{Introduction}
\IEEEPARstart{T}{\lowercase{he}} rapid advancement of modern information and communication technologies, particularly in the realms of the Internet of Things (IoT), beyond fifth-generation (B5G) and sixth-generation (6G) communication systems, and industrial IoTs, has driven a significant increase in the number of intelligent devices and a corresponding surge in the demand for intelligent applications. 
However, the computing capabilities of mobile devices are often insufficient to meet the growing demands for real-time data processing, low latency communication, and energy-efficient computation required by advanced IoT applications, B5G/6G networks, and industrial IoT scenarios. To address these challenges, mobile edge computing (MEC) has emerged as a promising solution, attracting significant interest from both academia and industry \cite{mec_1}.

MEC systems push computing power and storage resources to the network edge, significantly reducing data transmission latency and improving service quality. MEC servers are typically deployed at the network edge, close to mobile devices, allowing them to offload computation-intensive tasks to MEC servers for real-time computing services \cite{mec_2}.

To overcome the blockage of line-of-sight (LoS) links among an MEC server and the mobile users, and enhance the communication capacity among them, intelligent reflecting surface (IRS) technology, also known as reconfigurable intelligent surface, has been introduced \cite{irs_1}. In complex urban environments with buildings and tree obstructions, communication links among the MEC server and the mobile users are often non-line-of-sight (NLoS).
The study in \cite{irs_2} explores the use of IRS technology to mitigate the challenges associated with NLoS conditions in radar surveillance.
Specifically, the IRS consists of many controllable reflecting elements that can dynamically adjust the propagation path of wireless signals by tuning the amplitude and phase of these elements, thereby optimizing signal transmission quality and direction. By deploying the IRS to serve the MEC server and mobile devices, virtual LoS links can be established to improve signal quality and coverage \cite{irs_3, irs_4}. This results in more efficient and reliable data transmission for supporting high-density networks and mobile devices in urban environments \cite{irs_5}.

To better handle the massive multi-user interference anticipated in future B5G/6G networks and to further enhance the capacity of communication systems in dense mobile networks, rate-splitting multiple access (RSMA) technology has become a powerful multi-access method. The RSMA technology refers to a broad class of multi-user schemes that rely on the principle of rate splitting (RS). RS involves dividing user messages into two or more parts, allowing each part to be flexibly decoded at one or more receivers and managing interference among users \cite{rsma_1, Clerckx2023RSMA}. Through the RS approach, RSMA can unify the interference management strategies of non-orthogonal multiple access (NOMA) and space division multiple access, simultaneously applying full interference decoding and treating interference as noise.

{
The rest of this paper is organized as follows. 
Section \Rmnum{2} presents the related works and the motivations.
Section \Rmnum{3} introduces the IRS-assisted MEC system model and the problem formulation.
Section \Rmnum{4} proposes the hierarchical deep reinforcement learning (DRL) algorithm and presents the overall structure of the algorithm.
The numerical results are provided in Section \Rmnum{5}, and the paper is finally concluded in Section \Rmnum{6}.
}

\section{Related Works}
{Recent developments in MEC systems increasingly address dynamic resource coordination under complex network conditions.  While early approaches relied on centralized optimization frameworks, current trends shift toward distributed decision-making to handle real-time task scheduling in dense mobile environments \cite{8488481, 10443270, 9573404}.} 
\paragraph{MEC systems for distributed computing services}
Recently, several pioneering works have been proposed to implement MEC in B5G/6G systems. Specifically, computational task offloading is one of the core research directions in MEC \cite{mecrelate_1, mecrelate_2, mecrelate_3}.
Specifically, the optimization of computation task offloading with indivisible and delay-sensitive tasks was studied in \cite{mecrelate_1}, where a distributed offloading algorithm was designed to reduce average delay and task drop rate.
\cite{mecrelate_2} proposed a novel MEC-supported IoT network, combining communication and computation load balancing and resource allocation to reduce end-to-end delay in the integrated sensing, computation, and communication (ISCC) structure.
The joint task offloading and communication/computation resource allocation problem in a multi-user MEC system was studied \cite{mecrelate_3}, considering multi-channel scenarios and aiming to minimize the weighted sum of system energy consumption and latency.

Meanwhile,
the MEC technology has been integrated with some emerging communication scenarios \cite{mecrelate_4, mecrelate_5, 10554811}.
For instance, in \cite{mecrelate_4}, a space-air-ground integrated network was proposed for flexible MEC, allowing ground IoT devices to offload data packets to an unmanned aerial vehicle (UAV) and a low earth orbit (LEO) satellite for remote computation, thereby maximizing utility efficiency.
In \cite{mecrelate_5}, a maritime UAV-enabled MEC network for ISCC was studied, where the age of information is minimized by jointly optimizing the trajectory of the UAV, the total number of time slots, and the scheduling policy.
A UAV-enabled MEC system was studied in \cite{10554811}, where the energy consumption and latency considering fairness are minimized by optimizing the 3D trajectories of UAVs, transmission power of users, and the offloading policies while ensuring the security and fairness.
Moreover, a novel heterogeneous MEC system was studied in \cite{10606316}, where idle users are also tasked with computing for users experiencing a high volume of incoming tasks.

Although the previous works utilize enhanced transmission power or active beamforming at the transmitter to improve system throughput, there are still bottlenecks in channel capacity and energy efficiency.

\paragraph{IRS for enhanced communications with passive beamforming}
IRS has recently been applied in MEC systems to improve the network performance \cite{10093951, irsrelate_1, 10210080}.
For instance, the energy consumption of the hybrid access point (HAP) was minimized in an IRS-assisted MEC system \cite{10093951}, where the data offloading decision, the HAP’s active beamforming, and the IRS’s passive beamforming are jointly optimized.
In \cite{irsrelate_1}, the utility function, which consists of the energy efficiency, latency, and the price for user's data offloading, in an IRS-assisted MEC system was maximized.
{In \cite{PowerNetMax}, an architecture to optimize the IRS configuration and multi-IRS path was proposed for maximizing the received power, network reliability and system throughput in IRS assisted IoT networks.}
Besides, full-duplex IRS-assisted MEC system was studied in \cite{10210080}, where the dynamic and static IRS passive beamforming designs are joint considered to find a balance between the computational rate, system overhead as well as computational complexity of the method.

The integration of the IRS with other emerging technologies further substantiates their effectiveness in enhancing communication performance, improving signal quality, and optimizing resource allocation \cite{irsrelate_3, irsrelate_4, 9519632, irsrelate_2}.
For instance, an IRS-enabled single-objective multi-communication users integrated sensing and communications (ISAC) system was proposed in \cite{irsrelate_3}, and the transmit beamforming and reflecting beamforming were jointly designed.
Furthermore, the authors in \cite{irsrelate_4} studied a multi-IRS-enabled ISAC system.
In \cite{9519632}, an IRS-assisted ultra-reliable and low-latency communication (URLLC) system was investigated, where the total latency was minimized by jointly optimizing user grouping, blocklength allocation at the base station (BS), and passive beamforming at the IRS.
The authors in \cite{irsrelate_2} studied a multi-IRS-enabled vehicular edge computing network, a novel algorithm based on block coordinate descent and DRL was proposed to optimize data rate and cumulative energy utilization.
Recently, IRS-enabled MEC systems have shown remarkable potential in enhancing system efficiency and performance, particularly in terms of lower latency, higher data transmission rates, and improved energy efficiency.

Nevertheless, IRS-enabled MEC systems still face challenges such as managing the complex interactions among the configurations of the IRS and the computing resources of the MEC, as well as the inherent limitations of conventional multi-access techniques.
Moreover, traditional MEC systems often struggle with issues related to time latency and throughput, particularly in densely populated urban environment.
These limitations can lead to inefficient resource utilization and increased system delay.

\paragraph{RSMA for flexible decoding}

To overcome the limitations in traditional communication systems, such as low spectral efficiency, limited user access, inflexible resource allocation, and insufficient interference resistance, there is a growing interest in integrating advanced multiple access technologies, such as RSMA, into MEC systems to meet the demands for efficient, flexible, and large-scale user access in modern communication networks  \cite{rsmarelate_1, wang2024joint, rsmarelate_2}.
Several previous studies focused on the downlink transmission with RSMA.
Particularly, an energy efficiency maximization problem was investigated in \cite{rsmarelate_1} for IRS-assisted simultaneous wireless information and power transfer networks with RSMA leveraging the proximal policy optimization.
The system weighted energy efficiency was studied in \cite{wang2024joint} to fully utilize the spectrum and energy resources in a multi-IRS system with RSMA.
In \cite{rsmarelate_2}, RSMA was employed for the downlink transmission in a satellite communication system, where the power allocation was designed to maximize the total transmission rate.
Furthermore, \cite{rsmarelate_3, rsmarelate_4, rsmarelate_5} explored the RSMA applications in various scenarios, such as cognitive radio networks, massive multiple-input multiple-output cellular systems, and the UAV-enabled wireless networks.

{While most previous works have primarily focused on downlink RSMA transmission, uplink RSMA has received relatively less attention. However, uplink RSMA is equally important for improving the quality of service. Although some studies \cite{tegos2024distributeduplinkratesplitting} have explored uplink RSMA, further investigation is needed to fully exploit its potential, particularly in interference management.}
Therefore,
a few pioneering works \cite{rsmarelate_6, rsmarelate_7, rsmarelate_8} have investigated the uplink RSMA transmission.
In \cite{rsmarelate_6}, the optimization of decoding order and power allocation in an uplink system with RSMA was investigated to maximize the total throughput, and the performance gain was verified compared with other multiple access techniques such as NOMA.
In \cite{rsmarelate_7}, 
a multi-server MEC system integrated with RSMA was studied, and the maximum user delay was minimized in terms of the user concurrent task offloading.
The authors in \cite{rsmarelate_8} considered the total rate maximization in an IRS-assisted RSMA uplink system, and an efficient alternating optimization method combining successive convex approximation and Riemannian conjugate gradient was proposed to alternately optimize power allocation and passive beamforming of the IRS.

{Compared to other interference management strategies, such as NOMA and Space Division Multiple Access (SDMA), the primary advantage of RSMA lies in its ability to unify and adaptively apply both strategies by dynamically controlling the power allocation of the transmitted information. In downlink transmission scenarios, RSMA has been shown to inherently offer substantial benefits, including improved spectral efficiency, enhanced interference management, and better adaptability to user diversity \cite{rsmarelate_9}. However, in uplink transmission scenarios, RSMA faces a key challenge: the absence of an interference management strategy that fully capitalizes on its potential. This limitation hinders the effective utilization of resources, particularly in the integration of MEC systems with RSMA, resulting in a bottleneck that impedes the QoS for users in B5G/6G applications.}
{Table \ref{tab:rsma_relat_work} compares the details of the above RSMA related work and its interference management methods.}

\begin{table}[h!]
  \caption{RSMA Transmission Modes and Interference Management Approaches in related works}
  \label{tab:rsma_relat_work}
  \centering
  \begin{tabularx}{\linewidth}{>{\RaggedRight}X c >{\RaggedRight}X c} 
    \toprule
    \textbf{RSMA Transmission} & 
    \textbf{References} & 
    \textbf{Interference Management} \\
    \midrule
    RSMA downlink & 
    \cite{rsmarelate_1,wang2024joint,rsmarelate_2,rsmarelate_3} & 
    SIC + Treat interference as noise \\
    \addlinespace
    
    RSMA uplink & 
    \cite{rsmarelate_4,tegos2024distributeduplinkratesplitting,rsmarelate_6,rsmarelate_8} & 
    Full SIC \\
    \addlinespace
    
    RSMA uplink& 
    our work & 
    SIC + Treat interference as noise \\
    \bottomrule
  \end{tabularx}
  
  \vspace{-5pt}  % 微调caption间距
\end{table}

\subsection{Motivations and Contribution}
{The application of RSMA to uplink transmission introduces several challenges that need to be addressed. These include the limited receiver diversity (as only a single receiver is available in the uplink), the complexity involved in managing interference among multiple users, and the imperfect channel state information (CSI) that complicates the optimal rate splitting and decoding order. Additionally, resource allocation and scheduling for multiple users become increasingly complex, and integrating traditional interference management techniques, such as successive interference cancellation (SIC), with RSMA remains a significant challenge. While SIC can be effective in certain contexts, it does not fully exploit the dynamic interference management capabilities of RSMA in the uplink. 
To overcome these challenges, our work introduces a novel method for managing uplink interference within the RSMA framework. Essentially, we demonstrate how RSMA can outperform conventional methods in optimizing uplink interference management and resource utilization.} In the proposed algorithm framework, we focus on an IRS-assisted MEC system under the RSMA protocol, and jointly optimize the active beamforming of the BS, the passive beamforming of the IRS, the information splitting ratio and transmit power allocation ratio in RSMA, and MEC offloading to minimize the system delay.

Our research focuses on two key aspects. The first aspect is the design of the RSMA uplink transmission. Currently,
the existing works use traditional SIC technology to decode all sub-information in the RSMA uplink transmission.
However, this approach is not suitable in the uplink transmission compared to the traditional downlink transmission of the RSMA,
failing to highlight the significant advantages of the RSMA in interference management.
As a result, {we introduce a novel upstream data offload decoding method to the IRS-assisted MEC system to improve the throughput.} The second aspect is the design of the hierarchical DRL algorithm to solve the average time delay minimization problem.
Most studies utilize traditional DRL algorithms with fully connected networks (FCNs) to optimize in wireless communication systems.
While FCNs are well-suited for simple communication environments with a limited number of variables, they face challenges in more complex and high-dimensional scenarios. In these complex scenarios, the relationships between variables are not just simple one-to-one interactions; instead, they involve layered and dynamic dependencies that FCNs cannot model well. As a result, in high-dimensional environments, FCNs may not deliver the optimal performance needed for reliable communication systems \cite{irsrelate_5, ourvtc}.
Therefore, more computing resources are needed to support training in order to achieve better training performance.
To overcome the shortcoming of the FCNs, we propose a novel network structure integrated with the hierarchical DRL (HDRL) algorithm.
{The proposed HDRL algorithm increases the computational overhead on the MEC server side, but brings more refined interference management and lower latency for IRS-assisted MEC systems.}

The main contributions of this paper are summarized as follow:
\begin{itemize}
    \item We develop a novel IRS-assisted MEC system with RSMA, to minimize the system delay. The active beamforming of the BS, the passive beamforming of the IRS, the transmit power allocation, the ratios of public and private information allocation, the decoding order of offloading transmission, and the offloading ratios of tasks are jointly optimized.
    
    \item We propose a new RSMA uplink interference management method. Similar to the design in downlink RSMA, we split all information into two sub-information sets and apply interference-as-noise and interference-decoding techniques to the two sets, respectively. Our method can balance the two extreme strategies of fully decoding interference and treating interference as noise by adjusting the transmit power allocation of the two sub-information sets.
    
    \item We design a hierarchical DRL algorithm to optimize the discrete and continuous variables of the optimization problem. Considering the characteristics of the channel matrix, a novel network architecture combining CNN and DenseNet is proposed to extract features from the input channel information. Furthermore, this network architecture is applied across all neural networks within the algorithm.

    \item We conduct extensive experiments during training and testing processes, comparing the performance under different benchmark algorithms and communication methods.
    Numerical results verify that the proposed algorithm with the novel network structure achieves better performance than the benchmark algorithms.
\end{itemize}

\subsection{Notations}
The notation mentioned in this paper is introduced below.
$ {{\mathbb{C}}^{M\times N}} $ denotes the $ M \times N $ complex matrix.
For a generic vector $\boldsymbol{v} \in {\mathbb{C}}^{M}$, $ \Vert {\boldsymbol{v}}\Vert $ denotes the Euclidean norm, and $\mathrm{diag}(\boldsymbol{v}) \in {\mathbb{C}}^{M\times M}$ represents the $M \times M$ diagonal matrix with diagonal elements composed of $\boldsymbol{v}$ and all other elements being $0$.
$ {\cal C}{\cal N}(\mu ,{\sigma ^2}) $ denotes the circularly symmetric complex Gaussian distribution with $ \mu $ mean and $ {\sigma ^2} $ variance.
$ \mathrm{j} $ represents the imaginary unit, where $ {\mathrm{j}^2} = -1 $. For a generic matrix $ {\boldsymbol{G}} $, $ {{\boldsymbol{G}}^{\mathsf{H}}} $ and $ {{\boldsymbol{G}}^{\mathsf{T}}} $ denote the conjugate transpose and transpose of $ {\boldsymbol{G}} $, respectively.
And other symbols are summarized in Table \ref{tab:notations}.

% \paragraph{Organizations}
% The rest of this paper is organized as follows. The system model and problem formulation are introduced in Section II. Then, the hierarchical DRL algorithm for joint data offloading decision, the passive beamforming design of the IRS, the active beamforming design of the BS, and the decoding order of the RSMA protocol is presented in Section III. Afterwards, Section IV demonstrates the numerical results. Finally, Section V concludes this paper.

\begin{table*}[h!]
\normalsize 
	\caption{Summary of Notations Used in This Paper}
	\label{tab:notations}
    \renewcommand{\arraystretch}{1.5}
	\resizebox{\textwidth}{!}{ % 使用\resizebox来缩小表格
	\begin{tabular}{>{\raggedright\arraybackslash}m{1.5cm}|c|>{\raggedright\arraybackslash}m{1.5cm}|c}
 		\hline
		\textbf{Symbol} & \textbf{Definition} & \textbf{Symbol} & \textbf{Definition} \\
		\hline
		$M$ & Number of BS antennas & $N$ & Number of GUs\\ \hline
		$K$ & Number of IRS reflecting elements & $T$ & Total number of time slots \\ \hline
		$\tau$ & Duration of each time slot & $\Theta(t)$ & IRS reflection coefficient matrix at time slot $t$\\ \hline
		$\alpha_k$ & Amplitude reflection coefficient of the $k$-th IRS element & $\theta_k$ & Phase shift coefficient of the $k$-th IRS element\\ \hline
		$x_{n,c}(t)$ & Public sub-message of the $n$-th GU at time slot $t$ & $x_{n,p}(t)$ & Private sub-message of the $n$-th GU at time slot $t$ \\ \hline
		$p_n(t)$ & Transmit power of the $n$-th GU at time slot $t$ & $\gamma_n(t)$ & Power allocation ratio between public and private streams of the $n$-th GU at time slot $t$ \\ \hline
		$w_n(t)$ & The $n$-th beamforming vector at time slot $t$ & $\pi_n(t)$ & Decoding order of the $n$-th GU's public message at time slot $t$ \\ \hline
		$\beta_n(t)$ & Task offloading ratio of the $n$-th GU at time slot $t$ & $f_n$ & Local computing capability of the $n$-th GU \\ \hline
		$C_n$ & CPU cycles required to process one bit of task data at the GU $n$ & $\eta_n(t)$ & Ratio of public-to-private task allocation of the $n$-th GU at time slot $t$ \\ \hline
		$f_{\mathrm{MEC}}$ & Total computing frequency of the MEC server & $\varrho_n(t)$ & Computing frequency allocation ratio to the $n$-th GU at time slot $t$ \\ \hline
		$C_{\mathrm{MEC}}$ & CPU cycles required to process one bit at the MEC server & $f_c$ & Carrier frequency \\ \hline
	\end{tabular}
}
\end{table*}

%第二部分系统模型
\section{System Model and Problem Formulation}
\begin{figure}[h!]
	\centering
	\includegraphics[width=3.0in]{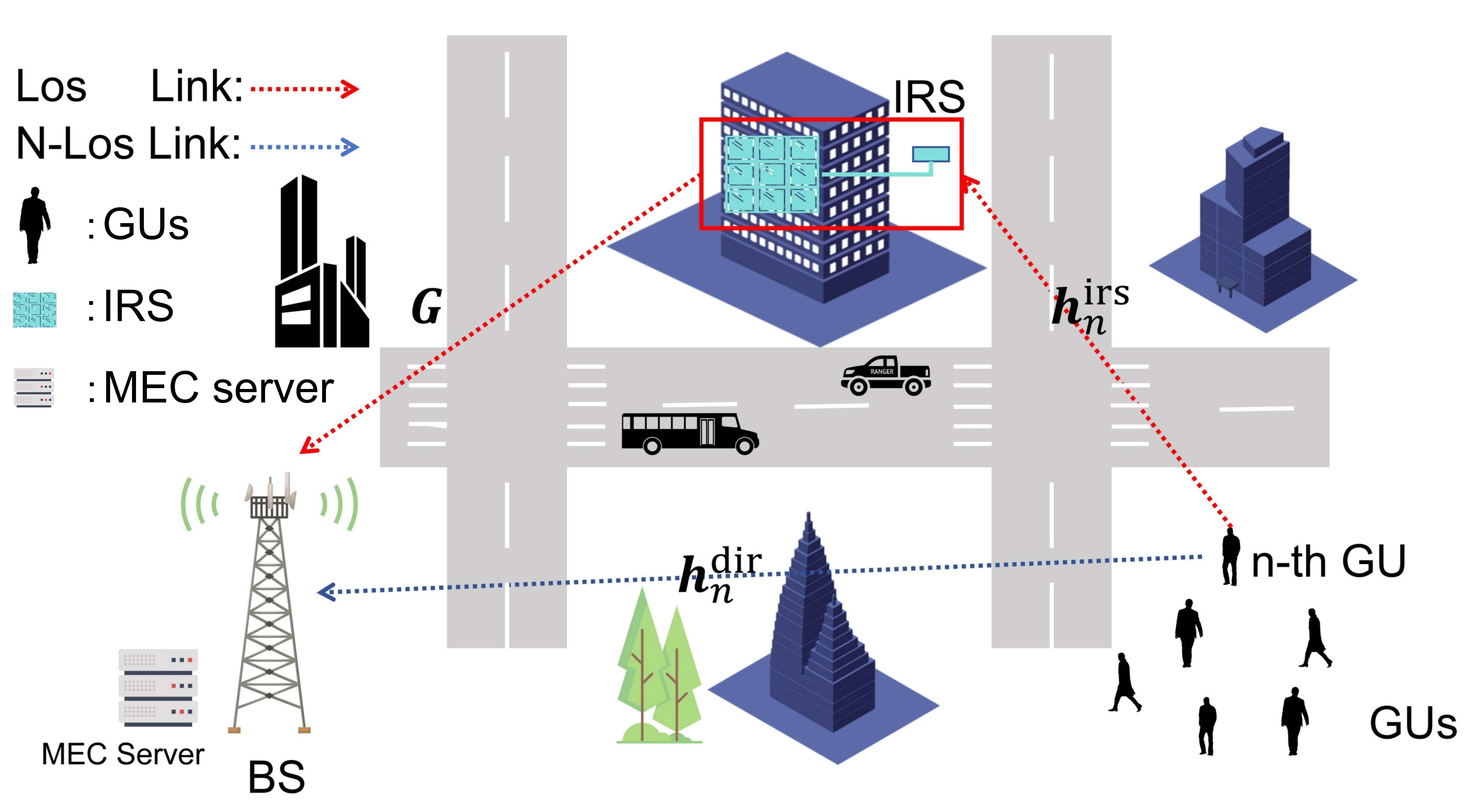}
	\caption{IRS-assisted MEC system.}
	\label{fig:system}
\end{figure}

As shown in Fig. \ref{fig:system},
we consider an IRS-assisted MEC system with RSMA technique. 
The system includes a BS with $M$ antennas, $N$ single-antenna mobile ground users (GUs),
an IRS with $K$ reflecting elements.
The BS is equipped with one MEC server.
The set of the GUs and the IRS elements are denoted as $\mathcal{N} =\left \{ 1,2,\ldots,N \right \} $ and $\mathcal{K} =\left \{ 1,2,\ldots,K \right \} $, respectively.
We consider a time interval with $T$ time slots, where the duration of each time slot is $\tau$ seconds.
The set of the time slots is denoted by a set $\mathcal{T} = \left \{ 1,2,\ldots, T \right \}$. It is assumed that at each time slot each GU has a task that needs to be computed within this time slot. For each time slot $t \in \mathcal{T}$, each user can decide to either process the task locally at the GU or partially offload the task to a MEC server for edge computation. Once the computation is completed in a given time slot, a new task will be generated at the beginning of the next time slot, continuing this cycle.

%子标题，无线通信模型
\subsection{Communication Model}
The reflection coefficient matrix of the IRS is defined as $\boldsymbol{\Theta(t)} = \text{diag}(\alpha_{1}e^{j\theta_{1}}, \alpha_{2}e^{j\theta_{2}}, \ldots, \alpha_{K}e^{j\theta_{K}})$, where $\alpha_{k} \in [0,1]$ represents the amplitude reflection coefficient, and $\theta_{k} \in [0, 2\pi)$ denotes the phase shift coefficient.
The elements of the IRS are typically designed to maximize the reflection gain and transmit the reflected signals towards the MEC server.
Similar to \cite{irs_alpha}, in practical applications, it is public to maximize the signal reflection of the IRS. Therefore, we set $\alpha_{k}=1$.

Let $\boldsymbol{h}_{n}^{\rm{dir}}(t) \in \mathbb{C}^{M \times 1}$ denote the equivalent baseband channel between the $n$-th GU and the BS at time slot $t$, $\boldsymbol{h}_{n}^{\rm{irs}}(t) \in \mathbb{C}^{K \times 1}$ denote the equivalent baseband channel between the $n$-th GU and the IRS at time slot $t$, and $\boldsymbol{G}(t) \in \mathbb{C}^{M \times K}$ denote the equivalent baseband channel between the IRS and the BS at time slot $t$.
Therefore, the composite channel gain from GU $n$ to the BS at time slot $t$ is given by
\begin{equation}
\label{信道增益}
\boldsymbol{H}_{n}\left ( t \right )  = \boldsymbol{G} \left ( t \right ) \boldsymbol\Theta \left ( t \right ) \boldsymbol{h}_{n}^{\rm{irs}} \left ( t \right ) +  \boldsymbol{h}_{n}^{\rm{dir}}\left ( t \right ) . 
\end{equation}

In the uplink RSMA system, each GU's information is split into $J$ sub-messages, which are transmitted to the BS via the IRS using non-orthogonal transmission. Without loss of generality, we consider that each GU's information $x_n(t)$ is divided into $J=2$ sub-messages, referred to as the public sub-message $x_{n,c}(t)$ and the private sub-message $x_{n,p}(t)$
\cite{rsmarelate_6}. Additionally, in an RSMA uplink system with $N$ users, splitting the messages of $K-1$ users is sufficient to avoid time-sharing and achieve every point in the capacity region \cite{Rimoldi1996RateSplitting}. For simplicity, we split all $N$ GUs' messages \cite{Clerckx2023RSMA}.

{Each GU has a maximum transmit power $P^{\max}_n$.
Let $\gamma_{n}(t)$ denote the power allocation ratio indicator that determines the power allocation for public and private transmission.
Specifically, $p_{n,c}(t) = \gamma_{n}(t) p_n(t)$ is the allocated power for public sub-message transmission, and $p_{n,p}(t) = (1 - \gamma_{n}(t)) p_n(t)$ represents the allocated power for private sub-message transmission.}
Thus, the message sent by the $n$-th GU at time slot $t$ can be expressed as
\begin{equation}
\label{用户发送的信息}
x_{n} \left ( t \right ) =  x_{n,c} \left ( t \right ) \sqrt{\gamma_{n}(t) p_n(t)}+  x_{n,p} \left ( t \right ) \sqrt{(1 - \gamma_{n}(t)) p_n(t)}.
\end{equation}

Hence, the received signal $\boldsymbol{y}\left ( t \right )\in \mathbb{C} ^{M\times 1 } $ at the BS can be expressed as Eq.(\ref{BS接受的信息}),
\begin{figure*}[!t]
\normalsize
\setcounter{mytempeqncnt}{\value{equation}}
\setcounter{equation}{2}
% \vspace*{-\baselineskip} % 取消额外的垂直间距
\begin{equation}
\label{BS接受的信息}
\begin{aligned}
\boldsymbol{y}\left ( t \right ) = \sum_{n=1}^{N} \boldsymbol{H}_{n}(t) x_n(t) + \boldsymbol{n}= \sum_{n=1}^{N} \boldsymbol{H}_{n}\left ( t \right )  \left \{   x_{n,c}\left ( t \right )\sqrt{\gamma_{n}(t) p_{n} \left ( t \right ) } + x_{n,p}\left ( t \right )\sqrt{(1-\gamma_{n}(t)) p_{n} \left ( t \right ) }\right \}+\boldsymbol{n}.
\end{aligned}
\end{equation}
\setcounter{equation}{\value{mytempeqncnt}}
\hrulefill
\end{figure*}where $\boldsymbol{n}=\left [ n_{1},n_{2}, \dots ,n_{M}  \right ]^{\mathsf{T}} $ is the additive white gaussian noise (AWGN).

The active beamforming vector at the BS is represented as $\boldsymbol{W}(t)\in \mathbb{C}^{M\times N }  $, where $\boldsymbol{w} _{n}(t) $ is the $n$-th column of the matrix $\boldsymbol{W}(t)$. Consequently, the recovered signal $\hat{x}_{n}(t)$ at the BS can be expressed as Eq.(\ref{恢复的信号}).
\begin{figure*}[!t]
\normalsize
\setcounter{mytempeqncnt}{\value{equation}}
\setcounter{equation}{3}
\vspace*{-\baselineskip} % 取消额外的垂直间距
\vspace*{1.5mm}
\begin{equation}
\label{恢复的信号}
\begin{aligned}
\hat{x}_{n}(t) =\boldsymbol{w}_{n} ^\mathrm{H}(t)\boldsymbol{y}\left ( t \right ) =\boldsymbol{w}_{n} ^\mathrm{H}(t)\left [\sum_{j=1}^{N} \boldsymbol{H}_{j}\left ( t \right ) \left \{   x_{j,c}\left ( t \right )\sqrt{\gamma_{j}(t) p_{j} \left ( t \right ) } + x_{j,p}\left ( t \right )\sqrt{(1-\gamma_{j}(t)) p_{j} \left ( t \right ) }\right \}+\boldsymbol{n}\right ].
\end{aligned}
\end{equation}
\vspace*{5mm}
\setcounter{equation}{\value{mytempeqncnt}}
\hrulefill
\end{figure*}
The conventional RSMA uplink decoding method utilizes SIC technology to decode both the public and private information of all users, without applying differentiated interference management strategies for different types of information. As a result, this approach does not fully leverage RSMA's advantages in interference management. To fully exploit the flexibility of the RSMA, an uplink interference management method at the receiver of the uplink RSMA system is designed in this paper. 
{In the proposed uplink RSMA scheme, each GU splits its information into a public part and a private part. At the MEC receiver, the BS first decodes the public information, treating all private information as noise. After removing the decoded public signals via SIC, the private information are then decoded.} Initially, public information is decoded and its interference is mitigated using SIC technology, followed by decoding the private information. For convenience, the decoding order of public information $x_{n,c} \left ( t \right ) $ from the $n$-th GU by the BS is defined as $\pi_{n}(t) \in \mathcal{S} =\left \{ 1,\ldots,S \right \} $, where $S=N$. Hence, the set of all decoding orders is defined as $\boldsymbol\Pi (t)=\left \{ \pi_{1}(t),\dots,\pi_{n}(t),\dots,\pi_{N}(t)  \right \} $.
If $\pi_{l}(t)>\pi_{n}(t)$, $l\ne n$, the signal of $x_{l,c}(t)$ is decoded after $x_{n,c}(t)$.

The signal-to-interference-plus-noise ratio (SINR) for decoding the information $x_{n,c} \left ( t \right ) $ and $x_{n,p} \left ( t \right ) $ are given by Eq.(\ref{SINR_public}) and Eq.(\ref{SINR_private}), respectively,
\begin{figure*}[!t]
\normalsize
\setcounter{mytempeqncnt}{\value{equation}}
\setcounter{equation}{4}
\vspace*{-\baselineskip} % 取消额外的垂直间距
\vspace*{-3.75mm}
\begin{equation}
\label{SINR_public}
\begin{aligned}
{\rho _{n,c} \left ( t \right )=\frac{\left |\boldsymbol{w}_{n} ^\mathrm{H}(t) \boldsymbol{H}_{n}\left ( t \right )   \right | ^{2} \gamma_{n}(t) p_{n} \left ( t \right )   }{ {\textstyle \sum_{\left \{ l, n\in \mathcal{N} |\pi _{l}(t)>  \pi _{n}(t) \right \} }} \left |\boldsymbol{w}_{n} ^\mathrm{H}(t) \boldsymbol{H}_{l}\left ( t \right )   \right | ^{2}\gamma_{l}(t) p_{l} \left ( t \right ) + \sum_{l=1}^{N}\left | \boldsymbol{w} _{n}^{\mathrm {H}}(t)\boldsymbol{H}_{l}(t)\right |^2(1-\gamma_{l}(t)) p_{l} \left ( t \right ) + \|\boldsymbol{w}_{n} ^\mathrm{H}(t) \|^2 \sigma ^{2}}}.
\end{aligned}
\end{equation}
\setcounter{equation}{\value{mytempeqncnt}}
\hrulefill
\end{figure*}
\begin{figure*}[!t]
\normalsize
\setcounter{mytempeqncnt}{\value{equation}}
\setcounter{equation}{5}
\vspace*{-\baselineskip} % 取消额外的垂直间距
\vspace*{1mm}
\begin{equation}
\label{SINR_private}
\begin{aligned}
{\rho _{n,p} \left ( t \right )=\frac{\left |\boldsymbol{w}_{n}^\mathrm{H}(t) \boldsymbol{H}_{n}\left ( t \right )   \right | ^{2}(1-\gamma_{n}(t)) p_{n} \left ( t \right )   }{\sum_{l=1,l\ne n}^{N}\left | \boldsymbol{w} _{n}^{\mathrm {H}}(t)\boldsymbol{H}_{l}(t)\right |^2(1-\gamma_{l}(t)) p_{l} \left ( t \right ) + \|\boldsymbol{w}_{n} ^\mathrm{H}(t) \|^2 \sigma ^{2}}}.
\end{aligned}
\end{equation}
\setcounter{equation}{\value{mytempeqncnt}}
\hrulefill
\end{figure*}
where $\sum_{\left \{ l, n\in \mathcal{N} |\pi _{l}(t)>  \pi _{n}(t) \right \} } \left |\boldsymbol{w}_{n} ^\mathrm{H}(t) \boldsymbol{H}_{l}\left ( t \right )   \right | ^{2} \gamma_{l}(t)) p_{l} \left ( t \right )$ represents the sum of all the subsequently decoded information and it is considered as interference. Besides, $\sigma ^{2}$ denotes the Gaussian noise power. Meanwhile, the denominators in Eq.(\ref{SINR_public}) and Eq.(\ref{SINR_private}) indicate that during the decoding of public information, all users' private information is treated as interference, and the public information of users decoded later is also regarded as interference.  When decoding private information, the private information of other users is similarly handled as interference. {By dynamically adjusting $\gamma_{n}(t)$, the mehod allows for flexible switching between these two decoding strategies, thereby optimizing interference management based on real-time channel conditions. The channel gain directly influences the choice of $\gamma_{n}(t)$, which in turn affects power allocation and decoding strategies. When the channel gain is sufficiently large, the RSMA decoding strategy fully shifts towards NOMA. In this scenario, the transmission power for private information is zero, while the transmission power for public information is one. Similarly, the information splitting ratio only contains public information. Conversely, when the channel gain is sufficiently small, the RSMA decoding strategy fully shifts towards SDMA.} 

Therefore, the achievable rate of the $n$-th GU for transmitting information $x_{n,c} \left ( t \right ) $ and $x_{n,p} \left ( t \right ) $ in time slot $t$ can be expressed as
\begin{equation}
\label{BS解码速率}
R_{n,c}\left(t\right)=B\log_{2}{\left(1+\rho _{n,c}\left(t\right )\right)},
\addtocounter{equation}{4}
\end{equation}
\begin{equation}
\label{BS解码速率}
R_{n,p} \left ( t \right ) =B\log_{2}{\left ( 1+ \rho _{n,p}\left ( t \right )   \right )},
\end{equation}
where $B$ is the system bandwidth.
%子标题，本地计算模型

\subsection{Local Computation Model}
At each time slot $t$, the $n$-th GU has a computation task with size $B_{n}\left ( t \right ) $, measured in bits. The GUs can offload part of the task data to the MEC server for computation or process it locally.
The task offloading ratio is defined as $\beta _{n} \left ( t \right ) \in \left [ 0,1 \right ] $, where $\beta _{n} \left ( t \right )=0$ means that the task is computed locally without any offloading, and $\beta _{n} \left ( t \right )=1$ denotes that the entire task is offloaded to the MEC server for computation.

Meanwhile, it is assumed that the task splitting procedure takes much less time compared to the transmission delay and the computation delay, and thus can be ignored \cite{downlinktask, downlinktask2, downlinktask3}.
Let $f_{n}$ denote the computing capability of the $n$-th GU,
which represents the CPU processing speed in CPU cycles per second.
Since the data volume of the computing result is usually much smaller than the original task volume, the time delay for transmitting the result can be ignored.

Let $C_{n} $ represent the number of CPU cycles required to process one bit of the task data. Therefore, the local computation delay can be expressed as
\begin{equation}
\label{本地计算延迟}
T_{n}^{\rm{local}}\left ( t \right ) =\frac{\left ( 1-\beta _{n}\left ( t \right )   \right )B_{n}\left ( t \right )C_{n}  }{f_{n} }  . 
\end{equation}

%子标题，任务卸载模型
\subsection{Task Offloading Model}
{Let $\eta _{n} \left ( t \right ) \in [0,1]$ denote the ratio of public-to-private task allocation of the $n$-th GU at time slot $t$. If $\eta _{n} \left ( t \right )=0$, there are only private tasks, while if $\eta _{n} \left ( t \right )=1$, there are only public tasks. Therefore, the data volumes of public and private information that need to be offloaded to the edge server for computation are defined as}
\begin{equation}
\label{公有数据大小}
B_{n}^{\rm{pub}} =\beta _{n} \left ( t \right ) B_{n} \left ( t \right ) \eta _{n} \left ( t \right ),
\end{equation}
\vspace{-1em}
\begin{equation}
\label{私有数据大小}
B_{n}^{\rm{pri}} =\beta _{n} \left ( t \right ) B_{n} \left ( t \right ) \left ( 1- \eta _{n} \left ( t \right )  \right ).
\end{equation}

Therefore, the time delay of the $n$-th GU for data offloading at time slot $t$ is written as
\begin{equation}
\label{数据卸载时间}
T_{n}^{\rm{trans}}\left ( t \right )  =\frac{B_{n}^{\rm{pub}}\left ( t \right )  }{R_{n,c}\left ( t \right )  } + \frac{B_{n}^{\rm{pri}}\left ( t \right )  }{R_{n,p}\left ( t \right )  }.
\end{equation}

We consider that the MEC server can perform task computation for different GUs simultaneously. The total computing frequency (CPU cycles per second) of the MEC server is denoted as $f_{\rm{MEC}} $. Let us denote the allocated computing frequency ratio to the $n$-th GU at time slot $t$ as $\varrho_n (t) \in (0,1)$, 
{
therefore, the $n$-th GU is assigned to a calculation frequency of  $\varrho_n(t) f_{MEC}$.
}
And the number of CPU cycles required to compute one bit data as $C_{\rm{MEC}} $.
Thus, the computation time for finishing the $n$-th GU task at the MEC server can be written as
\begin{equation}
\label{边缘服务器计算时延}
T_{n}^{\rm{MEC}} \left ( t \right ) =\frac{\beta _{n}\left ( t \right ) B_{n}\left ( t \right ) C_{\rm{MEC}}   }{\varrho_n(t)f_{\rm{MEC}} },
\end{equation}
where the sum of allocated computing resources cannot exceed the resource bound, i.e., $\sum_{n \in \mathcal{N}} \varrho_n (t) \leq 1$.

%子标题，问题表述
\subsection{Problem Formulation}
Since the task offloading and local computation of GUs are performed simultaneously, the total delay of the $n$-th GU at time slot $t$ is defined as the maximum of the offloading delay and the local computation delay, which is expressed as
\begin{equation}
\label{任务延迟}
T_{n} (t)=\max \left \{ T_{n}^{\rm{local}}\left ( t \right ),\left ( T_{n}^{\rm{trans}}\left ( t \right )+T_{n}^{\rm{MEC}}\left ( t \right ) \right )      \right \}  . 
\end{equation}

We aim to minimize the average delay by optimizing the passive beamforming of the IRS, active beamforming of the BS, the transmit power of the GUs, the data offloading allocation of the GUs, the decoding order of the RSMA, and the ratios of public and private information allocation of the GUs. Therefore, the optimization problem can be formulated as follows
\begin{subequations}\label{问题公式}
\begin{flalign}
  \textbf{P1}\ &\min_{\boldsymbol{W}, \boldsymbol\gamma ,\boldsymbol\beta , \boldsymbol\Pi  ,\boldsymbol\eta ,\boldsymbol\Theta } \ \ \frac{1}{T}\frac{1}{N}  \sum_{t\in \mathcal{T}}  \sum_{n\in \mathcal{N}} T_{n} \left ( t \right ) \tag{\ref{问题公式}} \\
 {\rm{s.t.}}  \quad & 0\le p_n (t) \le P_{n}^{\max}, \quad \forall n\in \mathcal{N},   \label{问题公式b}\\
& \beta _{n} \left ( t \right ) \in \left [ 0,1 \right ],  \quad \forall n\in \mathcal{N},\ \forall t\in \mathcal{T},\label{问题公式c}\\
& \eta _{n} \left ( t \right ) \in \left [ 0,1 \right ], \quad \forall n\in \mathcal{N},\ \forall t\in \mathcal{T}, \label{问题公式d}\\
& \gamma _{n} \left ( t \right ) \in \left [ 0,1 \right ], \quad \forall n\in \mathcal{N},\ \forall t\in \mathcal{T}, \label{问题公式e}\\
& \varrho_n \left ( t \right ) \in \left ( 0,1 \right ), \quad \forall n\in \mathcal{N},\ \forall t\in \mathcal{T}, \label{问题公式f}\\
& \sum_{n \in \mathcal{N}} \varrho_n \left ( t \right ) \leq 1,\ \forall t\in \mathcal{T}, \label{问题公式g}\\
& 0\le T_{n} \left ( t \right ) \le \tau,  \quad \forall n\in \mathcal{N},\ \forall t\in \mathcal{T},  \label{问题公式h}\\
& \theta _{k} \in \left [ 0,2\pi  \right ),  \quad \forall k\in \mathcal{K}, \label{问题公式i}\\
& \pi_{n}(t) \in \Pi (t),  \quad \forall n\in \mathcal{N},\ \forall t\in \mathcal{T},  \label{问题公式j}
\end{flalign}
\end{subequations}
where $\boldsymbol\gamma = \left \{ \gamma_{1}(t), \gamma_{2}(t), \ldots, \gamma_{N}(t) \right \}, \forall n, \forall t$ denotes the power allocation ratio, $\boldsymbol\beta = \left \{ \beta_{1}(t), \beta_{2}(t), \ldots, \beta_{N}(t) \right \}, \forall n, \forall t$ represents the offloading ratio, $\boldsymbol\eta = \left \{ \eta_{1}(t), \eta_{2}(t), \ldots, \eta_{N}(t) \right \}$, $\forall n, \forall t$ is the public-to-private ratio.
The constraint (\ref{问题公式b}) represents that the transmit power of GU-$n$ cannot exceed its maximal power threshold $P_n^{\max}$.
The constraints (\ref{问题公式c}), (\ref{问题公式d}) and (\ref{问题公式e}) denote the offloading ratio, the public-to-private ratio of the GUs, and the power allocation ratio, respectively.
The constraints (\ref{问题公式f}) and (\ref{问题公式g}) represent the MEC's computing frequency ratio allocation for GU $n$, and the sum of computing frequency ratio allocation for all GUs at the MEC server, respectively.
The constraint (\ref{问题公式h}) ensures that the computation tasks of the GUs must be completed within each time slot.
The constraint (\ref{问题公式i}) is the phase shift coefficient constraint of the reflect elements in the IRS.
Finally, the constraint (\ref{问题公式j}) denotes the decoding order.

\section{Hierarchical Learning for Task Offloading Strategy}

{The complexity of solving problem \textbf{P1} arises from three factors.  First, the optimization variables are interdependent.  For instance, the passive beamforming at the IRS affects the active beamforming at the BS, and the transmit power allocation for the GUs influences the data offloading decisions and decoding order.  These interdependencies create a highly coupled optimization problem.  Additionally, the discrete nature of constraint (\ref{问题公式j}), which relates to the phase shift coefficient of the IRS, introduces a combinatorial aspect to the problem. This makes it hard to solve the problem using conventional optimization methods.  Furthermore, the objective function is non-convex, which means that there are multiple local minima, and it is challenging to find the global optimum.}

To solve it, we propose a hierarchical DRL algorithm that combines deep Q-networks (DQN) and twin delayed deep deterministic policy gradient (TD3), which is referred to as convolutional denseNet enhanced HDRL (CDEH) algorithm. In this section, we present the overall structure of the algorithm and the design of the neural networks.

\subsection{Markov Design of Hierarchical DRL}
In this section, we introduce the elements of the Markov decision process (MDP) in our proposed CDEH algorithm.
{
At each time slot $t$, the proposed HDRL framework operates in a sequential manner, where continuous and discrete decisions are made in two coordinated stages:
Given the current CSI, the TD3 agent first determines the continuous control variables.
Based on the updated system configuration, the DQN agent selects the discrete decoding order for RSMA. 
Given both continuous and discrete decisions, the system computes the transmission rates, task completion delay, and reward. The agents are then updated using the shared reward signal.
}

\textbf{State:}
{The state space integrates CSI matrices from three links: GUs to BS, GUs to IRS, and IRS to BS. Since CSI matrices contain complex numbers, we decompose them into real and imaginary components. The state for both TD3 and DQN is defined as (\ref{状态空间}),}

\begin{figure*}[b] 
\hrule
\vspace{1em}
{
\begin{equation}
\label{状态空间}
\boldsymbol{s}(t) \in \left \{ \boldsymbol{h}^{\rm{dir}}_{\mathrm{imag}} (t),\boldsymbol{h}^{\rm{dir}}_{\mathrm{real}} (t),\boldsymbol{G}_{\mathrm{imag}} \left ( t \right ), \boldsymbol{G}_{\mathrm{real}} \left ( t \right ),\boldsymbol{h}^{\rm{irs}}_{\mathrm{imag}} \left ( t \right ),\boldsymbol{h}^{\rm{irs}}_{\mathrm{real}} \left ( t \right ) \right \},\ \forall t, 
\end{equation}
}
\end{figure*}
where $\boldsymbol{h}^{\rm{dir}}_{\mathrm{imag}} = \{\boldsymbol{h}^{\rm{dir}}_{1,\mathrm{imag}}, \boldsymbol{h}^{\rm{dir}}_{2,\mathrm{imag}}, \ldots, \boldsymbol{h}^{\rm{dir}}_{N,\mathrm{imag}} \}$, and $\boldsymbol{h}^{\rm{irs}}_{\mathrm{imag}} = \{\boldsymbol{h}^{\rm{irs}}_{1,\mathrm{imag}},\boldsymbol{h}^{\rm{irs}}_{2,\mathrm{imag}}, \ldots, \boldsymbol{h}^{\rm{irs}}_{N,\mathrm{imag}} \}$ denote the sets of imaginary parts of CSI from the GUs to the BS and from the GUs to the IRS, respectively. In the same way, $\boldsymbol{h}^{\rm{dir}}_{\mathrm{real}}$ stands for the real part.

\textbf{Action:}
Incorporating the optimization problem, the optimization variables are transformed into action decisions within the DRL algorithm. Specifically, task offloading allocation, RSMA power and task splitting ratio, passive beamforming at the IRS, and active beamforming at the BS are designed as continuous actions, while the decoding order is designed as a discrete action. TD3 is used for continuous action decisions, outputting task offloading allocation, RSMA power and task splitting ratio, passive beamforming of the IRS, and active beamforming at the BS. Therefore, the action space of TD3 is defined as follows:
\begin{equation}
\label{TD3动作空间}
\boldsymbol{a^{\text{\textmd{TD3}}}} (t) \in \left \{ \beta _{n} \left ( t \right ), \eta _{n} \left ( t \right ), \gamma _{n} \left ( t \right ), \boldsymbol{\Theta} (t), \boldsymbol{W}(t) \right \}.
\end{equation}

Meanwhile, DQN is utilized for designing actions in a discrete action space manner, where we only design the decoding order in a discrete form. In particular, DQN outputs the Q-values for all possible permutations of the decoding order, i.e., if there are $n$ public messages from $n$ users that need to be decoded sequentially, the output dimension is $n! $, and the decoding order is selected according to the $\epsilon$-greedy policy based on the corresponding Q-values. In our algorithm, DQN is responsible for making decoding order decisions. The action space of DQN is defined as follows:
\begin{equation}
\label{DQN动作a空间}
\boldsymbol{a^{\text{\textmd{DQN}}}} (t) \in \left \{ \Pi (t), \ \forall t \right \}.
\end{equation}

\textbf{Reward:}
When the agent takes action at time slot $t$, the reward $r(t)$ is calculated subsequently. Our objective is to minimize the total system delay. Since the DRL algorithm may adjust the action to maximize the reward value, the reward in our problem is defined as the opposite of the objective function of \eqref{问题公式}, which is given by
\begin{equation}
\label{奖励}
r(t) = - \frac{1}{T}\frac{1}{N} \sum_{t\in \mathcal{T}}  \sum_{n\in \mathcal{N}} T_{n} \left ( t \right ).
\end{equation}

{Since TD3 and DQN are optimized alternately, the overall reward is calculated after obtaining the action decisions from both algorithms.  This approach ensures that the system’s performance, measured by the single metric of average delay, is consistently optimized.  The average delay is directly influenced by the joint decisions made by TD3 and DQN.  Since these decisions collaboratively affect the data transmission rate, which in turn impacts the average delay, using a shared reward helps align the optimization objectives of both algorithms.} 
Therefore, the rewards for the TD3 algorithm, $\boldsymbol{r^{\text{\textmd{TD3}}}}(t)$, and the DQN algorithm, $\boldsymbol{r^{\text{\textmd{DQN}}}}(t)$, use the same reward, which is expressed as follows:
\begin{equation}
\label{DQN和TD3奖励}
\boldsymbol{r^{\text{\textmd{TD3}}}}(t) = \boldsymbol{r^{\text{\textmd{DQN}}}}(t) = r(t).
\end{equation}

{This unified reward facilitates effective collaboration between the two algorithms, ensuring a cohesive optimization process and improved overall system performance.}

\subsection{CNN DenseNet Empowered Neural Networks}

\begin{figure}[t]
	\centering
	\includegraphics[width=0.4\textwidth]{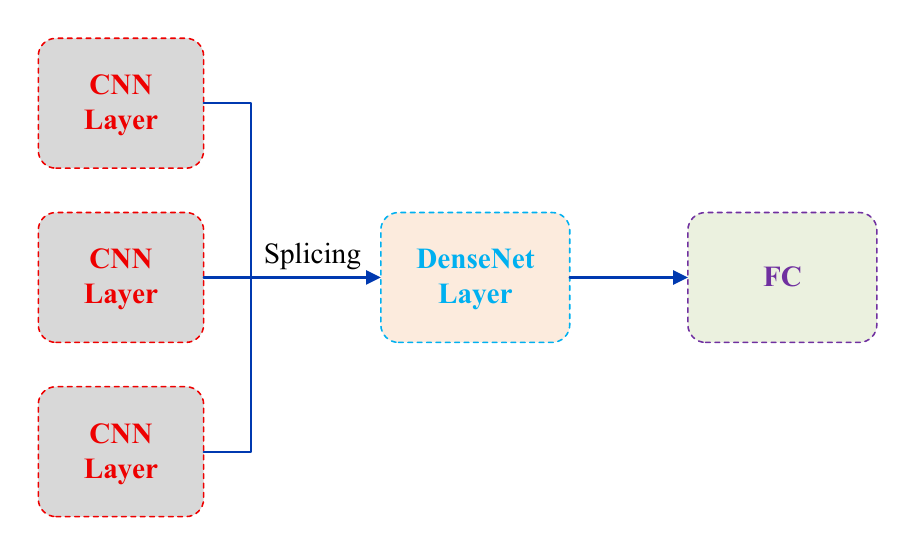}
	\caption{Overall Architecture and Flowchart of the Neural Network for TD3.}
	\label{fig:network_complete_flow}
\end{figure}

\begin{figure*}[t]
	\centering
	\includegraphics[width=0.7\textwidth]{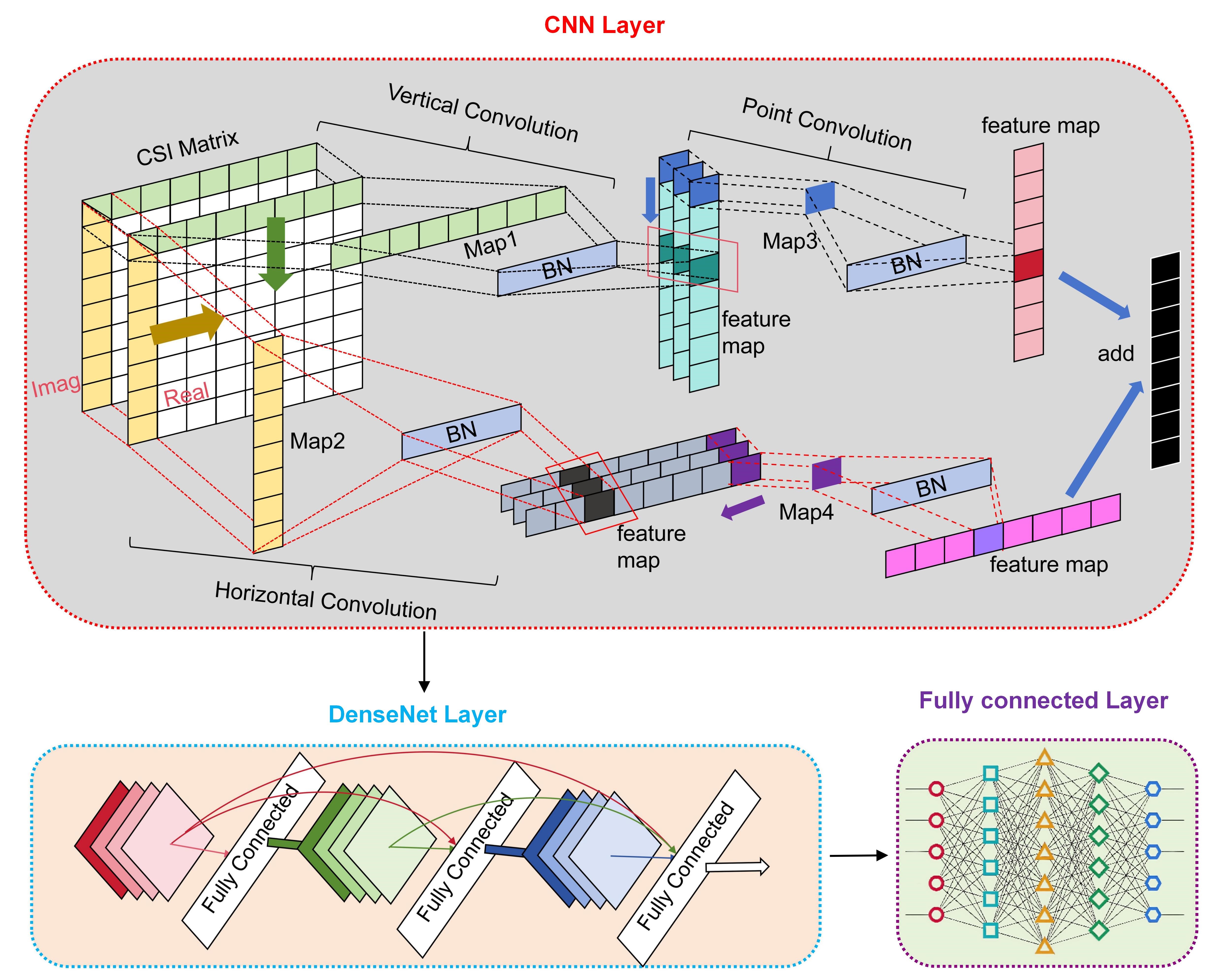}
	\caption{Detailed Flowchart of Neural Network Components for TD3.}
	\label{fig:network_struct}
\end{figure*}

{The complete neural network architecture of the proposed TD3 algorithm, along with the specific details of each network component, is illustrated in Fig \ref{fig:network_complete_flow} and \ref{fig:network_struct}. As shown in Fig \ref{fig:network_complete_flow}, the proposed neural network architecture of TD3, integrates multi-directional CNNs and DenseNet layers to capture spatial dependencies, followed by task-specific output layers. In this section, we will describe the overall data processing flow of the neural network and provide a detailed explanation of the structure and functionality of each part of the network.}

\subsubsection{Input Preprocessing: CSI Matrix Conversion}

{In communication environments, particularly in multi-antenna systems, matrix operations are frequently encountered.  When matrices are flattened and fed into a neural network, the inherent relationships between the rows and columns may be disrupted \cite{mishkin_2017}.  Additionally, since CSIs are complex numbers, simple extraction of the real and imaginary parts and concatenating them may lose valuable relationships, increasing training difficulty and instability.}

To preserve spatial relationships in complex-valued CSI matrices, we preserve the matrix form of the CSI matrices while separating the real and imaginary parts into two matrices of the same size, as follows:
\begin{align*}
&\boldsymbol{h}^{\rm{dir}}(t) \in \mathbb{C}^{M \times N} \rightarrow \boldsymbol{h}^{\rm{dir}}(t) \in \mathbb{R}^{M \times N \times 2}, \\
&\boldsymbol{h}^{\rm{irs}}(t) \in \mathbb{C}^{K \times N} \rightarrow \boldsymbol{h}^{\rm{irs}}(t) \in \mathbb{R}^{K \times N \times 2}, \\
&\boldsymbol{G}(t) \in \mathbb{C}^{M \times K} \rightarrow \boldsymbol{G}(t) \in \mathbb{R}^{M \times K \times 2}.
\end{align*}

{To extract features from each CSI matrix, three separate CNNs are used for the BS-IRS, BS-GUs, and IRS-GUs matrices.} 

\subsubsection{Feature Extraction: Multi-Directional CNNs}
{Each CNN branch processes a CSI matrix through two key operations: using $\boldsymbol{h}^{\rm{dir}}(t)$ as an example, vertical convolution with kernel size $1 \times N$ captures column-wise patterns (e.g., antenna-specific features), while horizontal convolution with kernel size $M \times 1$ extracts row-wise correlations (e.g., user-specific features).  Batch normalization (BN) is applied between convolutional layers.  Features from both directions are summed and flattened into a vector.  The outputs of all three CNNs are concatenated into a unified feature vector.}

\subsubsection{Feature Fusion: DenseNet Layers}
{The concatenated features are fed into three DenseNet layers to model cross-CSI dependencies.  DenseNet enhances the network's ability to capture complex spatial dependencies between CSI matrices by utilizing dense connections.  In this architecture, each layer receives feature maps from all previous layers, promoting feature reuse and improving gradient flow.  This design results in more stable training and improved performance by enriching the features passed to subsequent layers and aiding in better backpropagation.  Additionally, DenseNet's efficient structure enables the use of deeper networks without a significant increase in the number of parameters, making it ideal for processing high-dimensional CSI data.}

\subsubsection{Output Layers: Task-Specific Adaptation}
{The final layer of the network is customized for TD3 and DQN, with each layer designed to meet their specific requirements. In TD3, the architecture consists of the Actor-Critic framework. Both the Actor and Critic are implemented with three fully connected (FC) layers, mapping features to continuous actions and Q-values for state-action pairs.}

{The Q-network in DQN follows a similar structure to that in TD3, with some important differences due to the simpler decision-making process in DQN. Unlike TD3, DQN does not include the DenseNet section. Instead, it is composed only of the CNN and FC sections. The internal structure and the number of layers in both the CNN and FC components in DQN are identical to those in TD3.}

\subsection{CDEH algorithm}
In this work, we implement a hierarchical optimization strategy by alternately applying the TD3 and DQN algorithms. Initially, the state undergoes feature extraction and continuous action decision making through the TD3 algorithm. Concurrently, DQN handles discrete action decision making. The resulting action values drive interactions with the environment, yielding the subsequent state, the time delay for all GUs, and the corresponding rewards. These elements including the current state, subsequent state, actions, and rewards, are stored in the replay buffer as experience. Once the accumulated experience data meet the specified batch size, it is randomly sampled to update the neural networks. Notably, since the TD3 and DQN algorithms have their distinct natures in action decision-making, they employ different network update methods and structural configurations. The detailed descriptions are shown in the following.

\subsubsection{TD3}

The critic networks in TD3 are responsible for estimating the Q-values and are trained using a variant of the Bellman equation. TD3 uses two critic networks to mitigate overestimation bias. In TD3, there are four critic networks: two critic networks and two corresponding target networks. The Q-value update formula for the critic networks is
\begin{equation}
\label{td3的critic目标值}
y_i = r_i + \gamma \min_{j=1,2} Q_{\theta_j'}(s_{i+1}, \tilde{a}_{i+1}),
\end{equation}
where $Q_{\theta_j'}$ is the $j$-th target critic network, $\theta_j'$ is the parameter of the $j$-th target critic network, $i$ refers to the current time step, with $s_i$ representing the state, and $a_i$ representing the action at that time step, and $\min$ selects the smaller value from the two target Q-networks. The loss functions for training the two critic networks are
\begin{equation}
\label{td3的critic损失}
\mathcal{L}(\theta_j) = \frac{1}{M} \sum_{i=1}^{M} \left( Q_{\theta_j}(s_i, a_i) - y_i \right)^2, \quad \text{for } j = 1, 2,
\end{equation}
where $Q_{\theta_j}$ is the $j$-th critic network, $\theta_j$ is the parameter of the $j$-th critic network, and $M$ is the batch size, representing the total number of samples used in the loss computation. By minimizing these loss functions, the critic networks learn to approximate the Q-values accurately, enabling the actor network to improve its policy based on reliable Q-value estimation.

In TD3, the actor network, which represents the policy, is trained to maximize the expected return by taking actions that lead to higher Q-values. To improve training stability, the actor network's update is delayed relative to the critic networks. The loss function for the actor network is given by
\begin{equation}
\label{td3的actor损失}
\mathcal{L}(\psi) = -\frac{1}{M} \sum_{i=1}^{M} Q_{\theta_1}(s_i, \pi_{\psi}(s_i)),
\end{equation}
where $\psi$ is the parameter of the actor network, $Q_{\theta_1}$ is the first critic network, and $\pi_{\psi}(s_i)$ is the action predicted by the actor network for state $s_i$. The goal is to adjust the parameter $\psi$ to maximize the expected Q-value, which indirectly leads to maximizing the expected cumulative reward.

\subsubsection{DQN}
After the TD3 network has been updated, we proceed to update the DQN neural network. Both networks utilize the same sampled experiences for their updates. DQN employs a DNN to approximate the Q-value function, guiding the selection of the optimal action for a given state. The Q-value update rule in DQN is
\begin{equation}
\label{dqn目标Q值}
y_i = r_i + \gamma \max_{a_{i+1}} Q_{\phi'}(s_{i+1}, a_{i+1}),
\end{equation}
where $r_i$ is the immediate reward, $\gamma$ is the discount factor, and $Q_{\phi'}$ represents the target Q-network. The loss function for DQN is defined as
\begin{equation}
\label{dqn损失}
\mathcal{L}(\phi) = \frac{1}{M} \sum_{i=1}^{M} \left( Q_{\phi}(s_i, a_i) - y_i \right)^2,
\end{equation}
where $M$ is the minibatch size $Q_{\phi}$ represents the Q-network.

\begin{figure*}[t]
	\centering
	\includegraphics[width=6.5in]{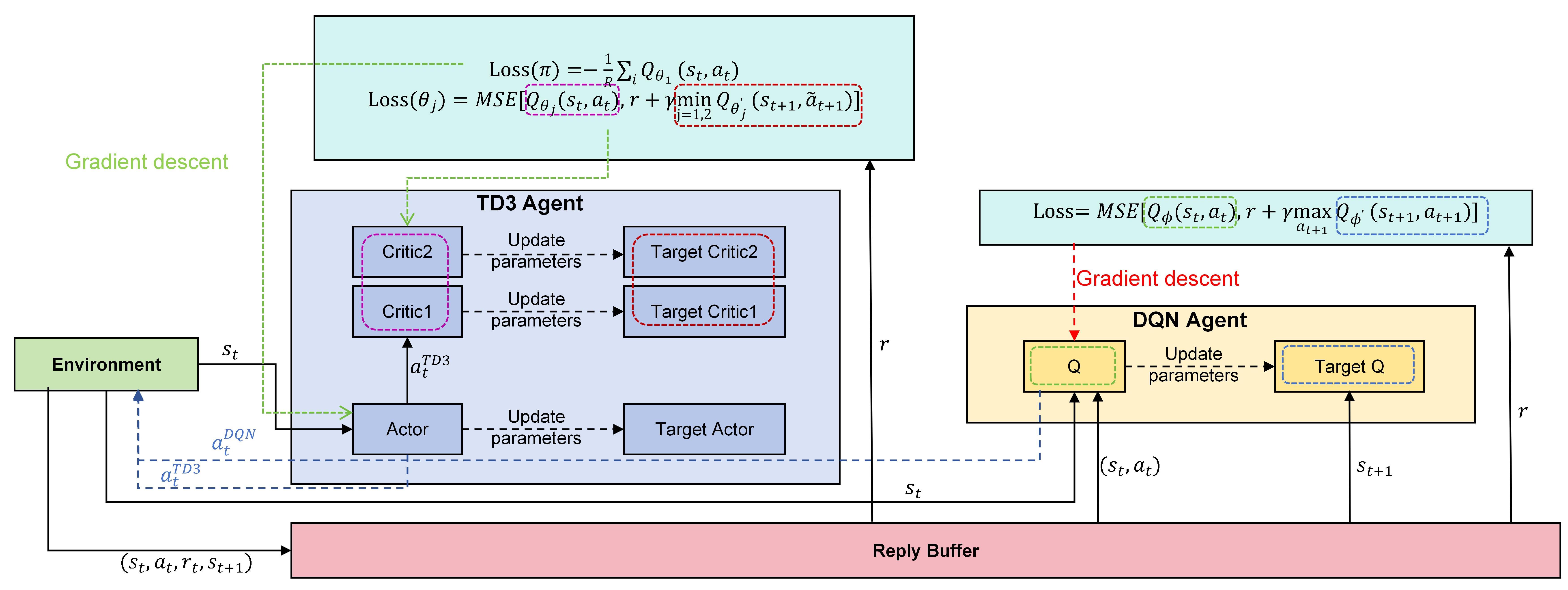}
	\caption{The workflow of proposed CDEH algorithm.}
	\label{fig:hdrl_system}
\end{figure*}

The workflow of the proposed CDEH algorithm on the hybrid DRL empowered by the modified structure is shown in Fig. \ref{fig:hdrl_system}. The details of the proposed CDEH algorithm is shown in Algorithm \ref{alg:CDEH}.

\begin{algorithm} \scriptsize
\caption{CDEH Algorithm}
\label{alg:CDEH}
\begin{algorithmic}[1] % 添加编号
\State \textbf{Input:} Overall CSI;
\State \textbf{Output:} The action chosen by the algorithm;

\State \textbf{Initialize:} 
\State \quad \textbf{TD3:} Critic networks $Q_{\theta_1}(s, a)$ and $Q_{\theta_2}(s, a)$ with parameters $\theta_1$ and $\theta_2$, actor network $\pi_{\psi}(s)$ with parameters $\psi$, target networks $Q_{\theta_1'}$ and $Q_{\theta_2'}$ with parameters $\theta_1' \leftarrow \theta_1$ and $\theta_2' \leftarrow \theta_2$, and target actor network $\pi_{\psi'}$ with parameters $\psi' \leftarrow \psi$;
\State \quad \textbf{DQN:} Q-network $Q_{\phi}(s, a)$ with parameters $\phi$, target Q-network $Q_{\phi'}$ with parameters $\phi' \leftarrow \phi$;
\State \quad Initialize shared replay buffer $\mathcal{R}$;

\For{epochs = 1 to $E_{\max}$ }
    \State Initialize environment and receive initial state $s_1$;
    \For{$t = 1, \ldots, T$}
        \State \textbf{TD3:}
        \State \quad Select action $a_t^{TD3} = \pi_{\psi}(s_t) + \epsilon$, where $\epsilon \sim \mathcal{N}(0, \sigma)$;
        
        \State \textbf{DQN:}
        \State \quad Select action $a_t^{DQN} = \arg\max_{a} Q_{\phi}(s_t, a)$ with $\epsilon$-greedy exploration;

        \State \textbf{Execute actions:}
        \State \quad Combine actions $a_t = \{a_t^{TD3}, a_t^{DQN}\}$;
        \State \quad Execute actions $a_t$ and observe reward $r_t$ and next state $s_{t+1}$;

        \State \quad Store transition $(s_t, a_t, r_t, s_{t+1})$ in $\mathcal{R}$;

        \State Sample a random mini-batch of $B$ transitions $(s_i, a_i, r_i, s_{i+1})$ from $\mathcal{R}$;

        \State \textbf{TD3:}
        \State \quad Compute target actions with added noise: $\tilde{a}_{i+1}^{TD3} = \pi_{\psi'}(s_{i+1}) + \epsilon$, where $\epsilon \sim \text{clip}(\mathcal{N}(0, \sigma_{\text{target}}), -c, c)$;
        \State \quad Compute target Q-value using (\ref{td3的critic目标值});
        \State \quad Update critic networks by minimizing the loss function (\ref{td3的critic损失});
        
        \If{$(t \mod d) == 0$}
            \State \quad Update actor network by maximizing the function (\ref{td3的actor损失});
            \State \quad Update target networks:
            $$
            \theta_j' \leftarrow \tau \theta_j + (1 - \tau) \theta_j', \quad j = 1, 2;
            $$
            $$
            \psi' \leftarrow \tau \psi + (1 - \tau) \psi';
            $$
        \EndIf

        \State \textbf{DQN:}
        \State \quad Compute target Q-value using (\ref{dqn目标Q值});
        \State \quad Update Q-network by minimizing the loss function (\ref{dqn损失});
        \State \quad Update target network:
        $$
        \phi' \leftarrow \tau \phi + (1 - \tau) \phi';
        $$

    \EndFor
\EndFor
\end{algorithmic}
\end{algorithm}

\subsection{Complexity Analysis}
The proposed CDEH algorithm is based on DQN and TD3 algorithms. For DRL algorithms, the complexity primarily arises from the training process of the neural network, including forward and backward propagation. The complexity of the first part of the network structure, which consists of a CNN, mainly stems from convolution operations and BN layer calculations. The CNN has four types of convolution kernels: $K\times1$, $N\times1$, $M\times1$, and $1\times1$. The number of input channel is 2, the number of intermediate output channel is 3, and the number of final output channel is 1. Therefore, the complexity of the first part of the network structure is $O(12(KM+KN+NM)+9D+4(M+N+K))$, where $D$ represents the output data size of the CNN. 
The term $12(KM + KN + NM)$ represents the computational cost of performing a convolution operation on the input channel matrix. Here, $KM$, $KN$, and $NM$ correspond to the dimensions of three types of channel information matrices. The factor $12$ arises from the fact that multiple convolutional kernels are applied across different channel matrices. There are four types of convolutional kernels: $K \times 1 \times 2$, $N \times 1 \times 2$, $M \times 1 \times 2$, and $1 \times 1 \times 3$, and each matrix has both real and imaginary components (hence the factor of $2$). Therefore, the complexity is multiplied by $12\ (2 \times 2 \times 3)$ to account for the different kernel sizes and channel dimensions. The expression $4(M + N + K)$ represents the number of times that the convolution kernels are computed with the input data. The factor $4$ represents the number of computations required for the third dimension (the channel), and $9D$ refers to the number of computations in the output layer, where $9$ corresponds to the computation count in the third dimension.

Subsequently, the complexity of the DenseNet part in the second section and the third section is $O(7D+4DH_1+H_1H_2+H_2|A|)$, where $H_1$ and $H_2$ represent the output sizes of the fully connected layers, $|A|$ represent the size of action space. Thus, the total complexity of the network is $O(12(KM+KN+NM)+16D+4(M+N+K)+4DH_1+H_1H_2+H_2|A|)$.

For the TD3 algorithm, there are six network structures: the actor network, target actor network, two critic networks, and two target critic networks. For DQN, there is only one Q network. All networks used in TD3 follow the same network structure, while the DQN network utilizes a three-layer FCNs. For simplicity, let $F$ represent the number of connections (computation steps) between all neurons in the DQN, $E$ denotes the number of training epochs, and $B$ represents the batch size. Therefore, the total complexity of the proposed CDEH algorithm is $O(6EB(12(KM+KN+NM)+16D+4(M+N+K))+4DH_1+H_1H_2+H_2|A|+6EBF)$.

\section{Numerical Results}
In this section, simulation results are presented to demonstrate the performance of the CDEH algorithm and communication design. The channel model combines large-scale and small-scale fading. Specifically, the direct channel between the GUs and the BS is modeled as the NLoS Rayleigh fading channel, while the IRS-reflected channels between the BS and IRS, and between the IRS and GUs, are modeled as LoS Rician fading channels. This approach captures the complex transmission characteristics in large-scale MIMO systems with IRS, considering both LoS and NLoS propagation environments. The IRS is located at a horizontal distance of 100m from the BS and at a height of 10m. The GUs are uniformly distributed within an annular region centered 50m from the IRS, with a radius ranging from $[2,10]$ meters. The channel vectors of GUs-BS, BS-IRS and IRS-GUs are modeled as

\begin{equation}
\label{dir_channel}
\boldsymbol{h}_n^{\rm{dir}} (t) = \sqrt{{\rm{PL}}_n^{\rm{dir}} (t)} \tilde{\boldsymbol{h}}_n^{\rm{dir}} (t),
\end{equation}

\begin{equation}
\begin{split}
\label{irs_user_channel}
\boldsymbol{h}_{n}^{\rm{irs}} \left ( t \right ) =\sqrt{{\rm{PL}}_n^{\rm{irs}} (t)} \Bigg(&\sqrt{\frac{\kappa}{1+\kappa}} \tilde{\boldsymbol{h}}_{n}^{\rm{irs, LoS}} (t) \\&
+\sqrt{\frac{1}{1+\kappa}} \tilde{\boldsymbol{h}}_{n}^{\rm{irs, N-LoS}} (t)\Bigg),
\end{split}
\end{equation}

\begin{equation}
\begin{split}
\label{bs_irs_channel}
\boldsymbol{G} \left ( t \right ) =\sqrt{{\rm{PL}}_{G} (t)}\Bigg(&\sqrt{\frac{\kappa}{1+\kappa}} \tilde{\boldsymbol{G}}^{\rm{LoS}}(t) \\&
+\sqrt{\frac{1}{1+\kappa}} \tilde{\boldsymbol{G}}^{\rm{N-LoS}} (t)\Bigg),
\end{split}
\end{equation}
where ${\rm{PL}}_n^{\rm{dir}}(t)$, ${\rm{PL}}_n^{\rm{irs}}(t)$, and ${\rm{PL}}_{G}(t)$ denote the large-scale fading in dB, and are given by

\begin{equation}
\begin{split}
{\rm{PL}}_n^{\rm{dir}}(t) = 20\log_{10}d_{n,{\rm{BS}}}(t) & + 20\log_{10}f_c + \\&
20\log_{10}\left(\frac{4\pi}{c}\right) + \ell_{\rm{N-LoS}},
\end{split}
\end{equation}

\begin{equation}
\begin{split}
{\rm{PL}}_n^{\rm{irs}}(t) = 20\log_{10}d_{n,{\rm{IRS}}}(t) & + 20\log_{10}f_c + \\&
20\log_{10}\left(\frac{4\pi}{c}\right) + \ell_{\rm{LoS}},
\end{split}
\end{equation}

\begin{equation}
\begin{split}
{\rm{PL}}_{G}(t) = 20\log_{10}d_{{\rm{IRS}},{\rm{BS}}} & + 20\log_{10}f_c + \\&
20\log_{10}\left(\frac{4\pi}{c}\right) + \ell_{\rm{LoS}}.
\end{split}
\end{equation}

Here, $d_{n,{\rm{BS}}}(t)$, $d_{n,{\rm{IRS}}}(t)$, and $d_{{\rm{IRS}},{\rm{BS}}}$ represent the distance from the $n$-th GU to the BS at time slot $t$, the distance from the $n$-th GU to the IRS at time slot $t$, and the distance from the IRS to the BS, respectively. In these equations, $f_c$ is the carrier frequency, $c$ is the speed of light, $\ell_{\rm{N-LoS}}$ represents the propagation loss of the N-LoS link, and $\ell_{\rm{LoS}}$ represents the propagation loss of the LoS link.

In (\ref{dir_channel}), (\ref{irs_user_channel}) and (\ref{bs_irs_channel}), the notation $\kappa$ is the Rician factor. The small-fading coefficient $\tilde{h}_n^{\rm{dir}} (t)$ is modeled as Rayleigh fading, i.e., $\tilde{h}_n^{\rm{dir}} (t) \sim \mathcal{C}\mathcal{N} (0,1)$, and $\tilde{\boldsymbol{h}}_{n}^{\rm{irs, LoS}} (t) = [1, e^{-j2\pi f_c d_0 \phi_{n}^{\rm{irs}} (t)}, \ldots, e^{-j2\pi f_c (K-1) d_0 \phi_{n}^{\rm{irs}} (t)}]^{\mathrm{T}}$ is the LoS steering vector. The N-LoS component $\tilde{\boldsymbol{h}}_{n}^{\rm{irs, N-LoS}} (t)$ follows the standard circularly symmetric complex Gaussian (CSCG) distribution \cite{irsrelate_5}.  $\tilde{\boldsymbol{G}}^{\rm{LoS}}(t) = [1, e^{-j2\pi f_c d_0 \phi_{\rm{IRS}}^{\rm{BS}}}, \ldots, e^{-j2\pi f_c (K-1) d_0 \phi_{\rm{IRS}}^{\rm{BS}}}]$ is the LoS steering vector, and N-LoS component $\tilde{\boldsymbol{G}}^{\rm{N-LoS}} (t)$ follows the standard CSCG distribution.
$d_0$ is the antenna separation distance, and $\phi_{n}^{\rm{irs}} (t) = \frac{q_{\rm{IRS}}^x - q_n^x(t)}{d_{n,{\rm{IRS}}}(t)}$ denotes the cosine of the angle of departure of the GU-to-IRS link.  $\phi_{\rm{IRS}}^{\rm{BS}} = \frac{q_{\rm{IRS}}^x - q_{\rm{BS}}^x}{d_{{\rm{IRS}},{\rm{BS}}}}$ represents the cosine of the angle of arrival of the IRS-to-BS link. 

Unless otherwise specified, the system parameters are set as follows: The number of antennas at the BS is $M=20$, the number of GUs is $N=5$, and the number of reflecting elements at the IRS is $K=50$. The task size for each GU is randomly generated within the range of $[0.4, 1.6]$ Kbits. The maximum transmission power $P^{\text{max}}_n$ is set to $5W$. The bandwidth is $B=400$ kHz, and the noise power is set to $\sigma^2 = -70$ dBm. {The other parameter settings of the neural network and CDEH algorithm are shown in Table \ref{tab:parameters}. Meanwhile, in terms of training strategy, TD3 utilizes delayed policy updates for the Actor network, soft updates, and target network techniques to enhance training stability and reduce overestimation bias.}

{The simulations in this work were conducted using one Tesla P100-PCIE GPU.  Each GPU is equipped with 16 GB of memory, which provides sufficient capacity for the computational demands of the proposed CDEH algorithm.  The driver version used is NVIDIA-SMI 460.39 with CUDA version 11.2.}

\renewcommand{\arraystretch}{1.5}
\begin{table*}[h!]
    \caption{{Network and CDEH algorithm Parameters Table}} % 将caption放在上方
    \centering
    \resizebox{\textwidth}{!}{ % 使用\resizebox来缩小表格
        \begin{tabular}{>{\raggedright\arraybackslash}m{6.5cm}|c|>{\raggedright\arraybackslash}m{6.5cm}|c}
        \hline
        \textbf{Parameter}          & \textbf{Value} & \textbf{Parameter}             & \textbf{Value} \\ \hline
        CNN Layer Number             &2
        & CNN Channels(Input, Middle, Output)   & 2, 3, 1
        \\ \hline
        CNN Output Dimension                & 64
        & DenseNet Layer Number      &3
        \\ \hline
        DenseNet Dimension   & 192, 192$\times$2, 192$\times$4            
        & Actor DenseNet Output Dimension      & 512 
        \\ \hline
        Critic DenseNet Output Dimension      & 1024
        & FC Neuron     & 512/1024, 256, 64     
        \\ \hline
        Soft Update             & 0.01      
        & $\gamma$              & 0            
        \\ \hline
        Actor Learning Rate          & 0.001        & Critic 1 Learning Rate     & 0.001        \\ \hline
        Critic 2 Learning Rate       & 0.0001       & Replay Buffer Size         & $2^{15}$     \\ \hline
        Batch Size                   & 256          & Noise Mean                 & 0.3          \\ \hline
        Noise Standard Deviation     & 0.2          & Noise Decay Rate           & 0.01         \\ \hline
        Policy Noise                 & 0.2          & Noise Clip                 & 0.5          \\ \hline
        Activation function          & ReLu Sigmod  & Optimizer                  & Adam 
        \\ \hline
        \end{tabular}
    }
    \label{tab:parameters}
\end{table*}

\subsection{Convergence}
\begin{figure}[htbp]
	\centering
	\includegraphics[width=3.5in]{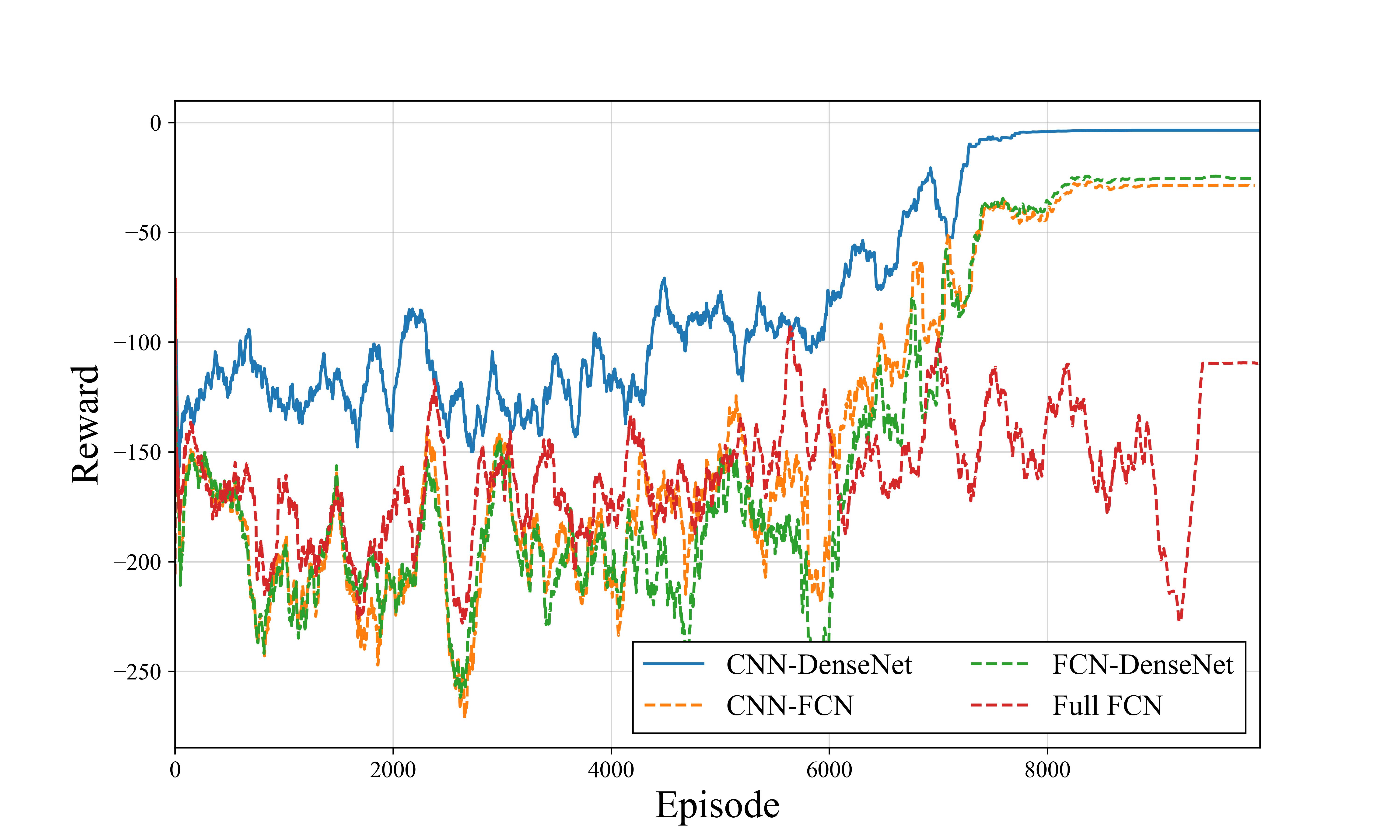}
	\caption{Training rewards for different network structures.}
	\label{fig:network_reward_curve}
\end{figure}

In Fig. \ref{fig:network_reward_curve}, we present a comparison of training trajectories based on the proposed network structure and other network combinations. {In FCN-DenseNet, we use FCN for feature extraction of the original CSI matrix and DenseNet for feature fusion. In the CNN-FCN, FCN is used to fuse the network CSI matrix features extracted by CNN. In Full FCN, feature extraction and feature fusion are done with FCN.} It is evident that the proposed CNN-DenseNet achieves better initial experiences, leading to higher reward values. When the time training reaches 7500 episodes, the algorithm nearly converges, fluctuating within a small range and achieving the highest reward among the four network combinations. The combinations of CNN-FCN and FCN-DenseNet achieve slightly lower reward values, displaying larger fluctuations throughout the training process. These two network architectures show instability, often requiring multiple training runs to achieve good results. For the training curve of the model Full FCN, the fluctuations are significant, indicating that it struggles to effectively learn the data features.  As a result, the reward after convergence at 8,500 episodes remains relatively low. This indicates that the CNN-DenseNet network structure effectively extracts channel characteristics, achieving better performance, while using either one alone also provides some performance improvement.

\begin{figure}[htbp]
	\centering
	\includegraphics[width=3.5in]{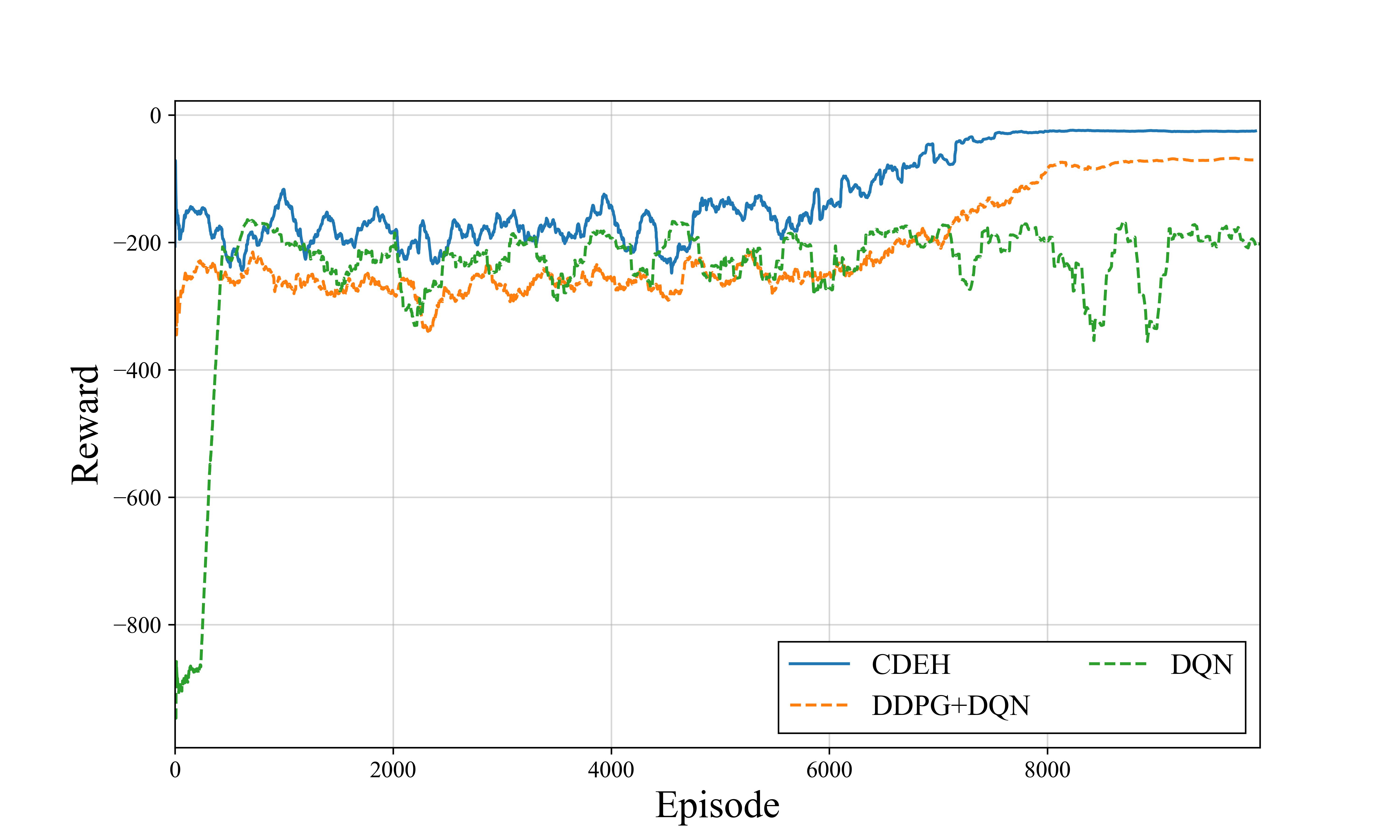}
	\caption{Training rewards for different algorithms.}
	\label{fig:dqn_ddpg_td3}
\end{figure}

In Fig. \ref{fig:dqn_ddpg_td3}, we compare the training curves under different DRL algorithms. 
{In DQN, all optimization variables are discretized. In DDPG + DQN, DDPG is used to optimize continuous variables, while DQN handles discrete decisions.} In the initial $2000$ episodes, DQN converges quickly, reaching performance slightly below the CDEH algorithm. This fast convergence is due to the discrete action space of DQN, which enables efficient decision-making. However, DQN suffers from limited exploration capabilities due to its reliance on a discrete set of actions. While the algorithm maintains its rewards within a certain range, the performance after convergence shows some instability and ultimately falls behind the CDEH algorithm. The final results indicate that the choice of discrete actions greatly affects DQN’s overall stability and performance. 

In contrast, the DDPG + DQN algorithm operates in a hybrid space, where DDPG optimizes continuous variables, and DQN handles the discrete decisions like decoding order.  While this combination allows for broader exploration in the continuous space, the initial episodes exhibit significant reward volatility due to the challenges of stabilizing training.  DDPG suffers from poor stability and is prone to local optima, which makes the training process less efficient.  Stabilization occurs between episodes $6,000$ and $8,000$.
The CDEH algorithm demonstrates consistently superior performance, balancing exploration and exploitation through a well-designed optimization process. Unlike the DQN algorithm, it avoids the limitations of discrete actions, and unlike the DDPG + DQN algorithm, it achieves stable training in a continuous space. As a result, the CDEH algorithm shows smooth convergence with minimal reward volatility, outperforming both the DQN and the DDPG + DQN by the end of training, around $8,000$ episodes. These results highlight that while the DQN algorithm excels in early convergence and the DDPG + DQN algorithm benefits from broad exploration, the CDEH algorithm effectively integrates the strengths of both approaches to achieve robust and consistent performance with superior stability.

\subsection{Impact of the decoding method }
\begin{figure}[htbp]
	\centering
	\includegraphics[width=3.0in]{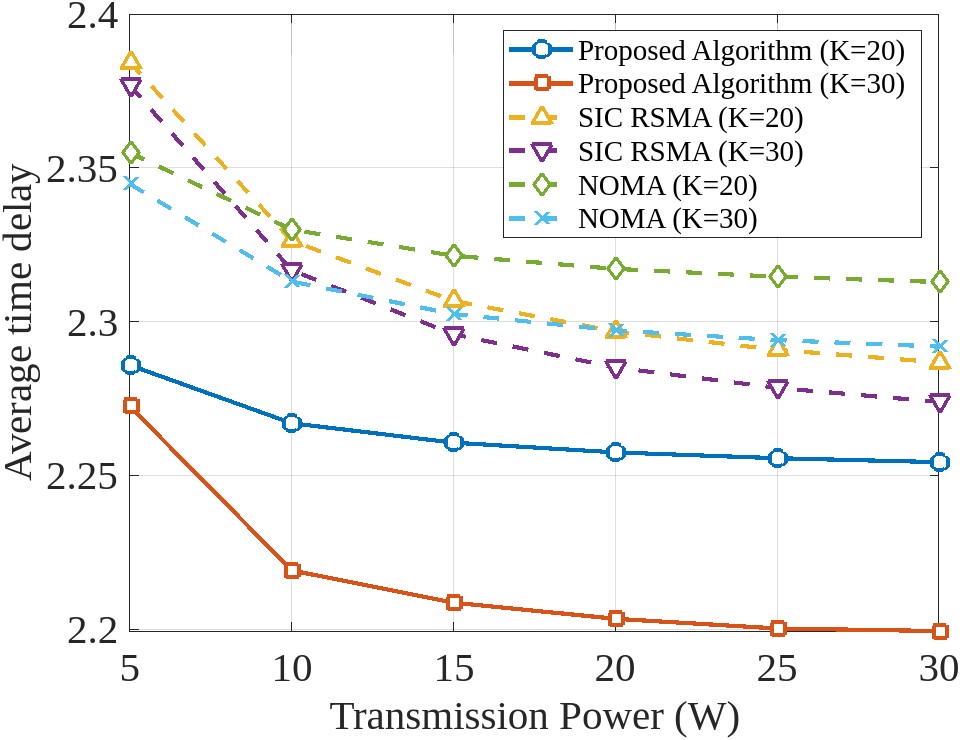}
	\caption{The average delay of different communication modes under different transmission power settings.}
	\label{fig:rsma_noma_sic_rsma}
\end{figure}

In our system model, we propose a new RSMA uplink communication method. As shown in Fig. \ref{fig:rsma_noma_sic_rsma}, we compare the average latency of different communication methods under varying $P^{\max}_n(W)$. It is evident that the RSMA strategy, implemented using the CDEH algorithm, outperforms other methods. Specifically, SIC RSMA is a widely-used approach for RSMA uplink transmission, where all information is split and SIC decoding is applied at the receiver for each information stream. It can be observed that when $P^{\max}_n$ is low, NOMA outperforms SIC RSMA.  This is because, at low power levels, the efficiency of SIC in RSMA is limited.  In RSMA, the power allocated to each information stream is split, and the decoding relies heavily on accurate interference cancellation.  However, with low transmission power, the SINR for each decoded stream is reduced, making the cancellation process less effective.  As a result, RSMA's average latency performance lags behind NOMA, which does not split streams and thus maintains stronger individual signal strengths.
As $P^{\max}_n$ increases, the performance of SIC RSMA gradually surpasses NOMA.  The increased power allows RSMA to allocate power more flexibly across streams, improving the SINR and enhancing the effectiveness of interference cancellation.  This results in significant performance improvements, with RSMA ultimately outperforming NOMA at higher power levels. Additionally, it can be observed that the proposed RSMA strategy allows for more flexible power allocation and interference management strategies, further enhancing performance based on SIC RSMA. This demonstrates the superiority of the proposed flexible uplink interference management method for RSMA.

\subsection{Impact of the number of IRS elements}
\begin{figure}[htbp]
	\centering
	\includegraphics[width=3.0in]{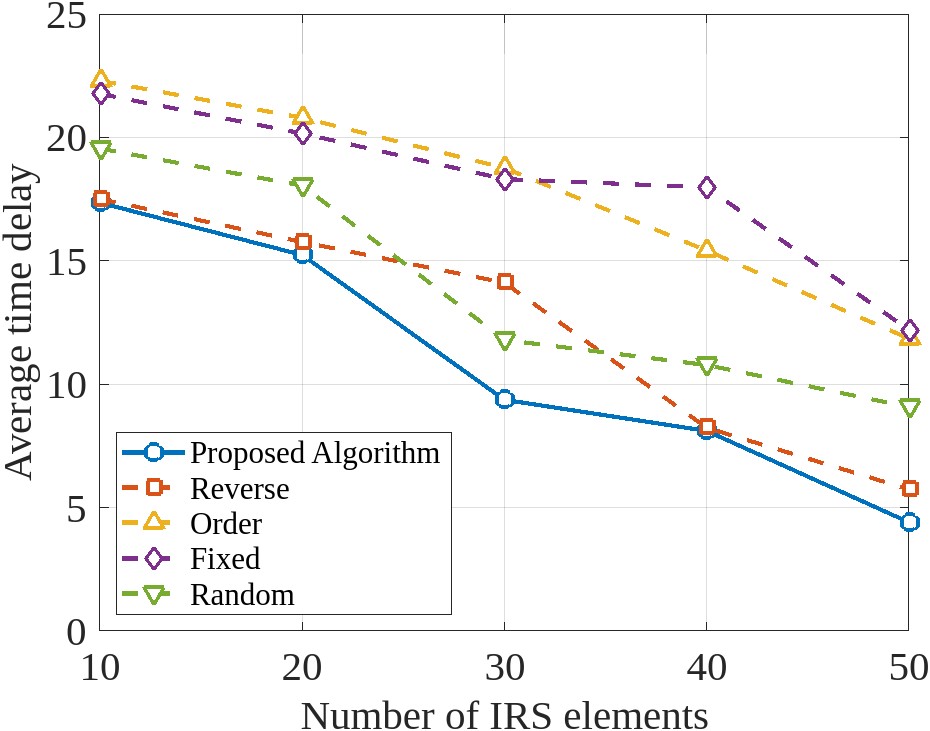}
	\caption{Performance comparison of algorithms with different decoding orders under different numbers of IRS elements.}
	\label{fig:irs_num_decode_order}
\end{figure}

\begin{figure}[htbp]
	\centering
	\includegraphics[width=3.0in]{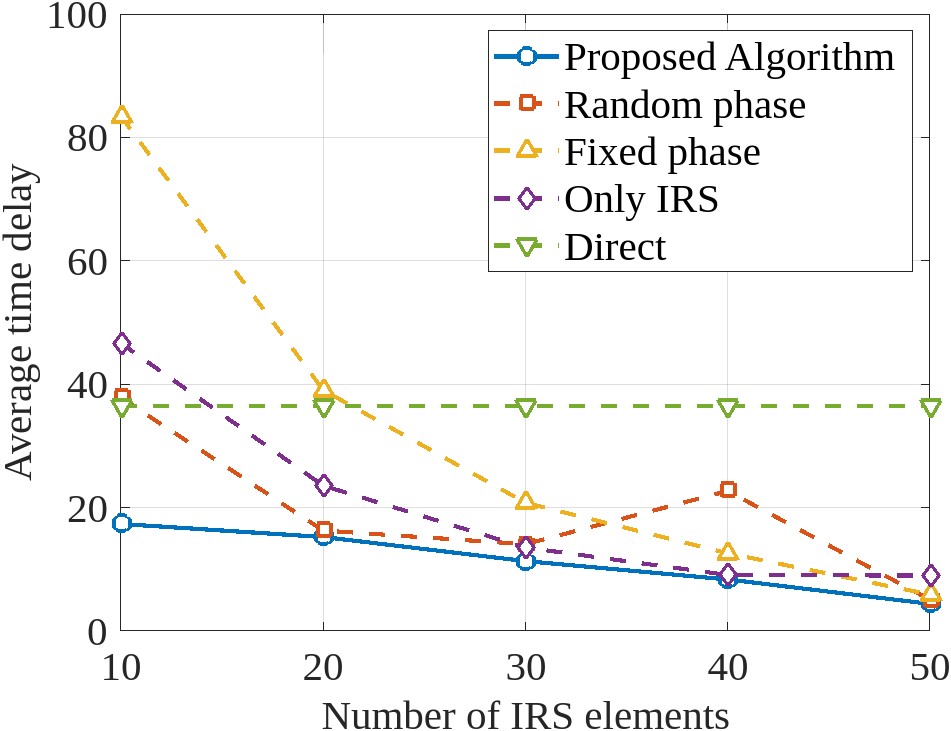}
	\caption{Performance comparison of algorithms with different IRS phase shifts under different numbers of IRS elements.}
	\label{fig:irs_num_irs_phase}
\end{figure}

\begin{figure}[htbp]
	\centering
	\includegraphics[width=3.0in]{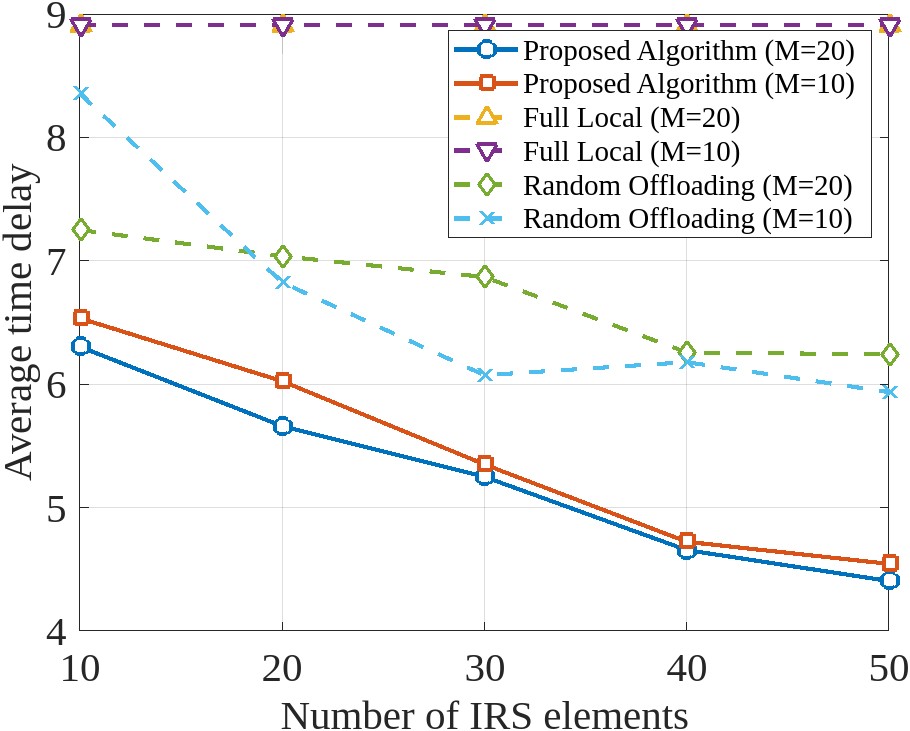}
	\caption{Performance comparison of algorithms with different offloading strategies under different numbers of IRS elements.}
	\label{fig:irs_num_offload_decision}
\end{figure}

In Figs. \ref{fig:irs_num_decode_order}, \ref{fig:irs_num_irs_phase} and \ref{fig:irs_num_offload_decision}, we compare the average  time delay performance of the CDEH algorithm and baseline schemes under different numbers of IRS elements. Specifically, as the number of IRS elements increases, the communication environment's gain from IRS progressively enhances, resulting in a gradual reduction in average time delay. Moreover, it is evident that our proposed CDEH algorithm consistently achieves much lower average delays compared to the other benchmarks.

In Fig. \ref{fig:irs_num_decode_order}, we compare the latency performance of four different decoding algorithms: reverse decode, proposed algorithm, order decode, and fixed decode. These algorithms differ primarily in their decoding order strategies. 
{The reverse decode algorithm (Reverse) decodes signals in the reverse order based on the user sequence (e.g., from user $N$ to user $1$), while the order decode algorithm (Order) follows a fixed sequential order (e.g., from user $1$ to user $N$). The fixed decode algorithm (Fixed) maintains a pre-determined decoding order that does not adapt to changes in network conditions, whereas the Proposed Algorithm, based on a DQN approach, dynamically adjusts the decoding order to optimize performance based on the network state.}
In Fig. \ref{fig:irs_num_decode_order}, it can be observed that the Reverse decode algorithm and the Proposed Algorithm achieve lower latency compared to the Order decode and Fixed decode algorithms. The performance of the Reverse decode algorithm is close to that of the Proposed Algorithm under some numbers of IRS elements. This is because the optimal decoding order of the received signals learned by the DQN algorithm is similar to the reverse decode order, usually only two or three signals have different decoding orders. 

As the number of IRS elements increases, the latency for all algorithms decreases.  This trend occurs because increasing the number of IRS elements improves the quality of the communication links by providing more opportunities for reflective signal paths, thereby enhancing the overall channel conditions.  Improved channel conditions result in stronger received signals, higher signal-to-noise ratio (SNR), and more efficient interference management.  Consequently, the decoding algorithms can operate more effectively, leading to a reduction in transmission latency. When the number of IRS elements exceeds 30, the latency of the fixed decode algorithm becomes greater than that of the  order decode algorithm. When the number of IRS elements increases to 40, the latency reduction for the fixed decode algorithm is minimal. 

In Fig. \ref{fig:irs_num_irs_phase}, the overall trend indicates that the proposed algorithm consistently achieves the lowest average latency compared to other benchmark algorithms as the number of IRS elements increases. This demonstrates the effectiveness of the proposed algorithm in efficiently leveraging IRS gains. 
{The four algorithms being compared in this figure include: Random phase, where the IRS phase shifts are randomly selected; Fixed phase, where the IRS phase is fixed and does not change dynamically over time; Only IRS, where only the IRS-reflected channel is considered, without direct transmission; Direct, where only the direct channel is considered without IRS assistance.} It can be observed that the latency for Direct does not decrease with an increasing number of IRS elements, as Direct does not account for IRS gains, rendering the number of IRS elements irrelevant to performance. The performance of Fixed phase and Only IRS is worse than Direct when the number of IRS elements is small, indicating that the gain from IRS is not significant at low IRS unit numbers.   However, as the number of IRS elements increases and the gain becomes more pronounced, both Fixed phase and Only IRS achieve lower average latency than Direct. When the number of IRS elements increases to 50, the performance of Fixed phase surpasses that of Only IRS. This indicates that even though increasing the number of IRS elements does not affect Direct and the performance of Direct is not optimal, a combination of reflection and direct communication link performs better than Only IRS when the number of IRS elements is sufficiently large. This is evident from the consistently lower average latency of the proposed algorithm. The proposed algorithm delivers the best performance, particularly as the number of IRS elements increases. It ensures steady and consistent reductions in latency as IRS elements are added, outperforming the benchmark methods by efficiently optimizing both the reflection and direct communication paths. The uniform reduction in latency with the increasing number of IRS elements emphasizes the robustness of the proposed algorithm in IRS-enhanced communication systems\footnote{It is worth noting that in our experiments, when the number of IRS reflection units becomes excessively large, such as 100 or more, the action space of the DRL algorithm increases significantly. This results in less effective optimization and a plateau in performance improvement. Consequently, we did not explore whether increasing the number of IRS elements would continue to improve performance steadily or if it would eventually reach a point of diminishing returns. However, based on the performance trends observed with the Fixed Phase method, it is reasonable to anticipate a saturation point, beyond which the addition of more IRS reflection units yields negligible performance gains. This suggests that while the impact of IRS elements on performance remains an area worth exploring, diminishing returns are likely to occur after a certain number of IRS elements are introduced.}.

In Fig. \ref{fig:irs_num_offload_decision}, we compare the effects of different offloading strategies and the number of antennas on the average latency under varying numbers of IRS elements. {The Full Local algorithm performs all computations locally, and the Random Offloading algorithm randomly allocates the ratio between the local computations and the task offloading to the MEC server.} It can be seen that the latency of the Full Local algorithm is unaffected by the number of IRS elements. Among the algorithms, the Full Local algorithm performs all computation locally without any offloading. This explains why its latency remains constant regardless of the number of IRS elements. Since no offloading is involved, the quality of the wireless transmission link, enhanced by the IRS elements, does not affect its performance.

On the other hand, the CDEH algorithm achieves the best performance under varying IRS elements. This is because the CDEH algorithm can effectively utilize IRS and beamforming, adapting its strategy based on the network conditions. As the number of IRS elements increases, the communication link improves, which facilitates a higher data transmission rate to the MEC, thereby reducing the overall latency. Additionally, it can be observed that an increase in the number of antennas helps reduce average latency across all algorithms. More antennas contribute to better beamforming and signal strength, leading to an improved SINR and more efficient data transmission. For CDEH algorithm, more antennas enable better exploitation of the IRS elements and improve the communication quality, further reducing latency.

\subsection{Impact of the transmission power}
\begin{figure}[htbp]
	\centering
	\includegraphics[width=3.0in]{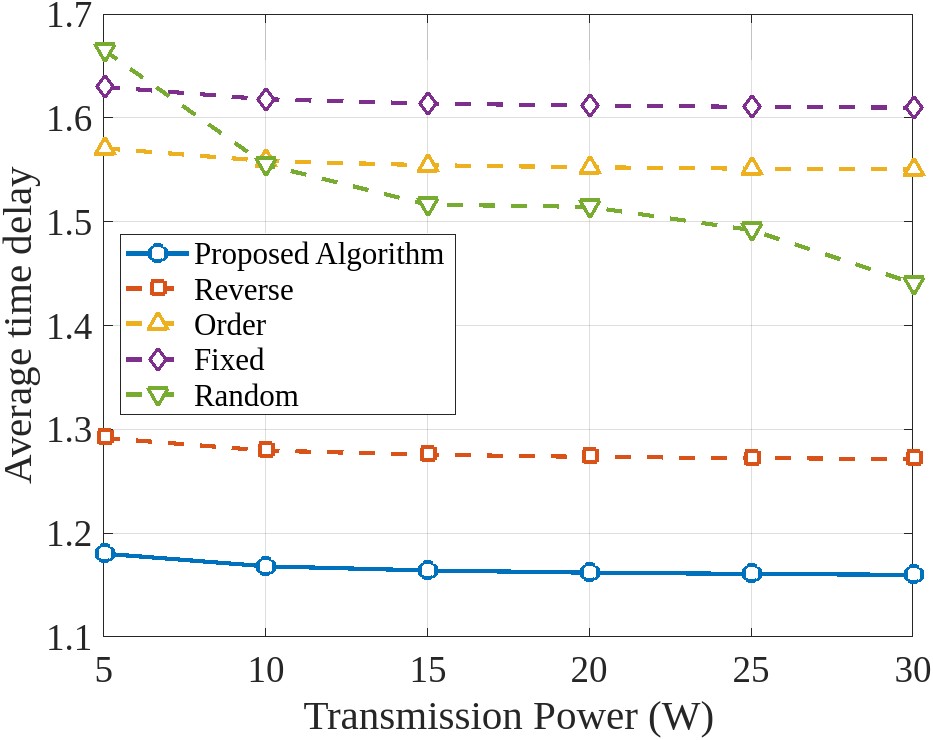}
	\caption{Performance comparison of algorithms with different decoding orders under different transmission power thresholds.}
	\label{fig:power_decode_order}
\end{figure}

\begin{figure}[htbp]
	\centering
	\includegraphics[width=3.0in]{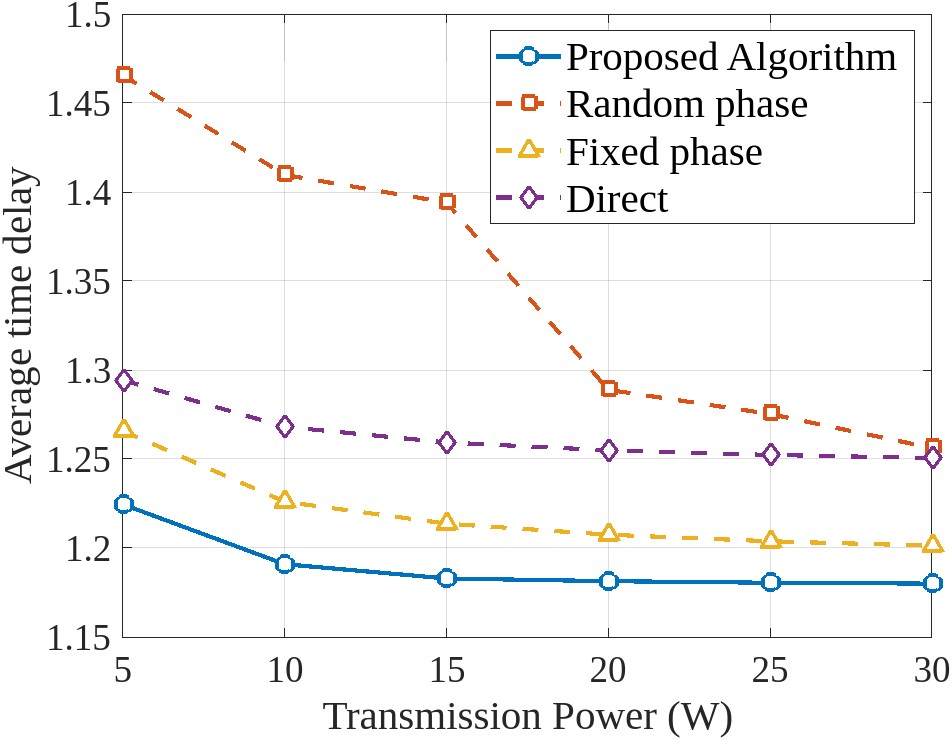}
	\caption{Performance comparison of algorithms with different IRS phase shifts under different transmission power thresholds.}
	\label{fig:power_irs_phase}
\end{figure}

\begin{figure}[htbp]
	\centering
	\includegraphics[width=3.0in]{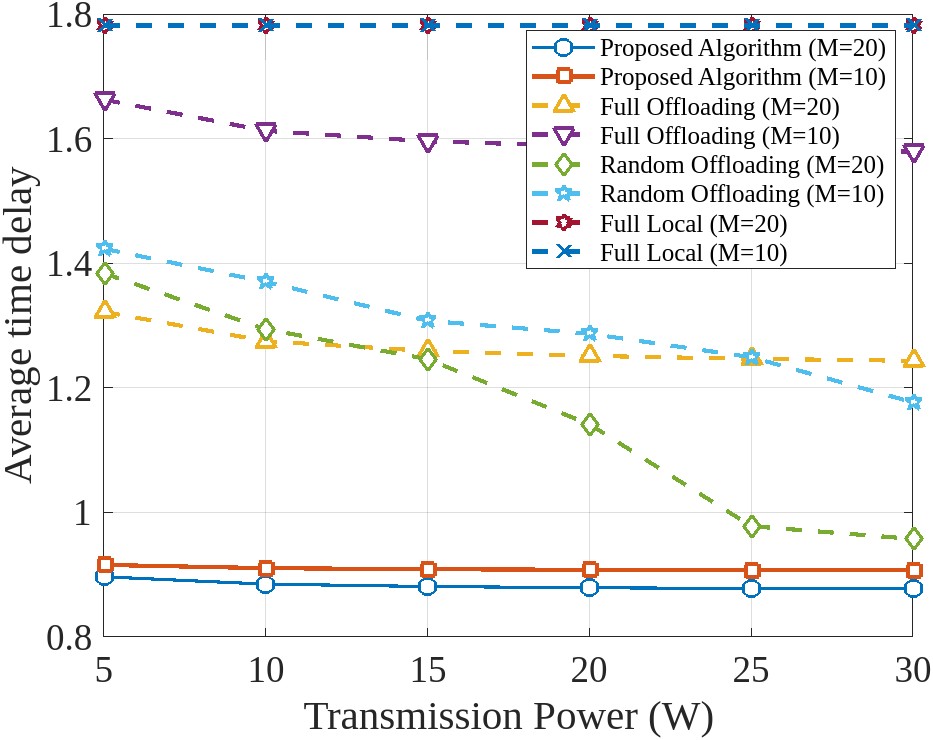}
	\caption{Performance comparison of algorithms with different offloading strategies under different transmission power thresholds.}
	\label{fig:power_offload_decision}
\end{figure}

In Figs. \ref{fig:power_decode_order}, \ref{fig:power_irs_phase} and \ref{fig:power_offload_decision}, we present the impact of different maximum transmission power, $P^{\max}_n$, on the average latency for our CDEH algorithm and benchmark schemes. Meanwhile, the number of IRS reflecting elements is set to $50$. Specifically, as $P^{\max}_n$ increases, the average system latency decreases gradually. The reason for this reduction in latency is that increasing the maximum transmission power enhances the quality of the communication link by boosting the SNR. With a higher transmission power, the transmitted signals can overcome interference and noise more effectively, leading to faster and more reliable data transmission.
However, in some communication environments, this results in significant performance improvements, while in others, the improvement is less pronounced. The CDEH algorithm consistently achieves the lowest average latency.

In Fig. \ref{fig:power_decode_order}, it is evident that the proposed CDEH algorithm has the lowest average latency, followed by the Reverse decode, Order decode, Fixed decode, and Random decode algorithms. Unlike the effect of the number of IRS elements on performance, increasing transmission power does not cause performance crossover between different decoding orders. This is because increasing the number of IRS elements adds additional reflection channels, generating more state information for DRL input, making the neural network sensitive to the number of IRS elements. Additionally, the increase in IRS elements significantly impacts phase optimization and the calculation of final channel gains. However, transmission power is not a part of the DRL state input and does not affect state information. The impact of $P^{\max}_n$ mainly lies in the calculation of communication rates. Since the transmission power is split and allocated to two sub-information, the system can dynamically adjust the transmission power of the sub-informations by adjusting the power allocation ratio. Consequently, the impact of $P^{\max}_n$ on the system is weakened. Meanwhile, the optimization results of the decoding order also reduce the impact of $P^{\max}_n$, making the performance curve of $P^{\max}_n$ smoother than that of the number of IRS elements.

In Fig. \ref{fig:power_irs_phase}, we observe as transmission power $P^{\max}_n$ increases, the system's average latency achieved by different IRS phase shift schemes smoothly decreases. As $P^{\max}_n$ increases, the average latency becomes flatter. This indicates that larger transmission power results in lower average latency, but as the transmission power increases beyond a certain point, the impact of $P^{\max}_n$ on performance diminishes. 
This is because, while higher transmission power continues to improve the SNR and transmission quality, other factors such as network congestion and computational bottlenecks at the edge servers start to limit the overall system performance. As a result, beyond a certain power threshold, further increases in $P^{\max}_n$ yield smaller latency gains, demonstrating a reduced influence of transmission power on overall latency.
Furthermore, the CDEH algorithm is observed to achieve the best performance among all the IRS phase optimization schemes, demonstrating its effectiveness. This highlights the importance of optimizing IRS phase shifts to enhance system performance, as an suboptimal or random phase configuration can lead to suboptimal outcomes.

In Fig. \ref{fig:power_offload_decision}, the local computation curve remains a horizontal line, since $P^{\max}_n$ only affects the communication transmission rates and does not impact the computation time. The local computation is constrained by the processing capability of the GU, which remains constant regardless of the transmission power. Therefore, increasing $P^{\max}_n$ does not contribute to any reduction in the local computation time.
On the other hand, both Full Offloading and the Proposed Algorithm show performance improvement as the number of BS antennas increases. This is because having more antennas enhances the array gain, leading to improved beamforming capabilities. With better beamforming, the SNR increases, allowing for faster data transmission. Consequently, this reduces the overall latency by accelerating the offloading process and enhancing the efficiency of data transmission between the user device and the BS. This benefit is particularly pronounced in the CDEH algorithm, where both the number of antennas and optimized resource allocation strategies work together to minimize latency.

\subsection{Impact of the number of users}
\begin{figure}[htbp]
	\centering
	\includegraphics[width=3.0in]{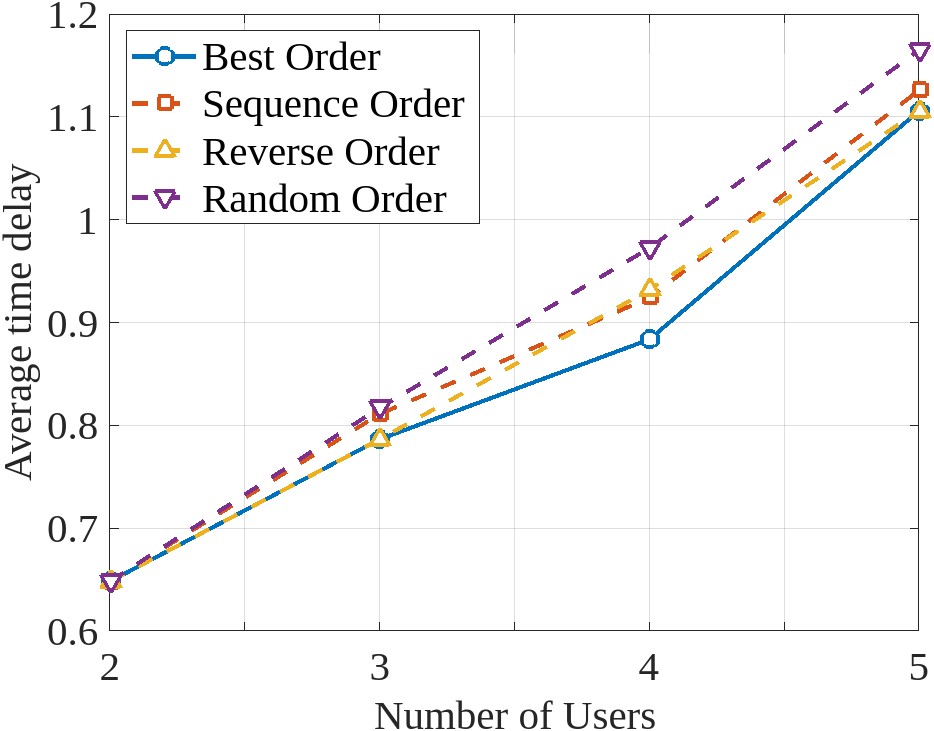}
	\caption{Performance comparison of algorithms with  different decoding orders under different numbers of users.}
	\label{fig:user_num_decode}
\end{figure}

\begin{figure}[htbp]
	\centering
	\includegraphics[width=3.0in]{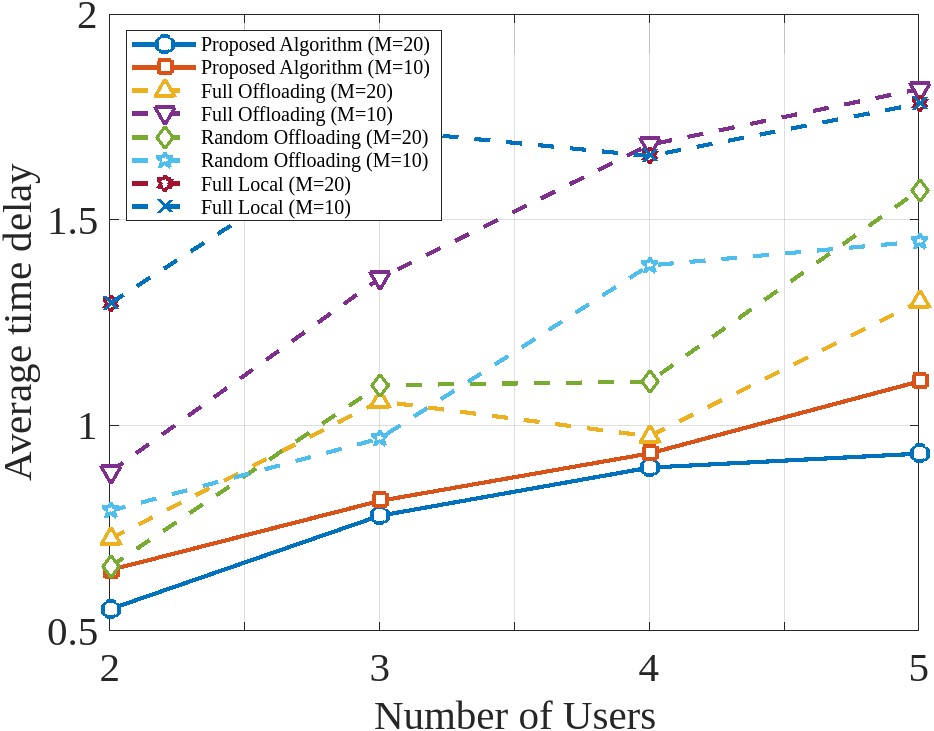}
	\caption{Performance comparison of algorithms with different offloading strategies under different numbers of users.}
	\label{fig:user_num_offload_decision}
\end{figure}

In Figs. \ref{fig:user_num_decode} and \ref{fig:user_num_offload_decision}, we compare the impact of different optimization variable settings on average time latency under varying numbers of users. In Fig. \ref{fig:user_num_decode}, we compare the effect of different decoding order settings on latency. 
As expected, the total latency gradually increases as the number of users grows.  This is because, with more users, the system has to handle a larger number of tasks in each time slot, leading to higher overall computation and communication demands. Additionally, the increase in latency can be attributed to the fact that network resources, such as bandwidth and computational power, are fixed and shared among users.  As the number of users increases, each user receives a smaller portion of resources, which in turn slows down both the data transmission and processing speed.  This results in longer delays, as more users must compete for the available resources, thereby increasing the overall system latency.

When the number of users is small, such as $N=2$ and $N=3$, the number of signals that need to be decoded is small, thus the effect of the decoding orders on  the time latency is not significant, resulting in minimal or no performance difference between different decoding orders. For instance, when $N=2$, the performance of different decoding orders is nearly identical. 
However, as the number of users increases, the performance gap between different decoding order schemes widens.  This is because, with more users, the system must handle a larger number of decoding tasks simultaneously, making the order in which these tasks are processed more critical to overall system efficiency.

In Fig. \ref{fig:user_num_offload_decision}, we compare the average time latency of various offloading strategies under different antenna counts and offloading decision as the number of users increases. Overall, it can be observed that as $N$ increases, the latency for all algorithms tends to increase due to the higher computational and communication demands. However, the proposed algorithm consistently outperforms other strategies, maintaining the lowest latency across different user counts and system configurations. This highlights the efficiency of the proposed approach in managing resource allocation and task scheduling.

When $N=4$, $M=20$, the average latency for Full Offloading decrease. This phenomenon occurs because our goal is to minimize the average latency. As the number of users increases from $N=3$ to $N=4$, the task size generated by the additional user is smaller than the average task size when $N=3$. In some cases, the new user's task size might even be smaller than the minimum task size in the previous scenario, which causes a reduction in the average latency. This effect can be attributed to the system’s ability to dynamically allocate resources more efficiently when task sizes are unevenly distributed. When the new task requires less resources, it reduces the overall load on the system, leading to a temporary improvement in performance. However, as the number of users continues to grow beyond $N=4$, the average latency starts to increase again due to the system being more heavily loaded. This is evident in Full Local, where local computation is unaffected by the communication environment. Meanwhile, it is clear that increasing the number of antennas has a beneficial effect on system performance for the proposed algorithm.

\section{Conclusion}
In this paper, we have proposed a hierarchical DRL-based CDEH algorithm to minimize the average delay in an IRS-assisted MEC system with RSMA.
The active beamforming of the BS, the passive beamforming of the IRS, the transmit power allocation, the task splitting allocation, the decoding order, and the offloading allocation are jointly optimized.
To address the challenges of RSMA in the uplink transmission, we have introduced a novel interference management strategy.
Furthermore, to enhance the agent's ability to extract valuable information from the CSI matrix, we have designed a network structure that combines CNN and DenseNet.
Numerical results indicate that the proposed method and network structure outperform traditional fully connected networks and RSMA uplink strategies. The combination of neural networks effectively extracts channel features, resulting in better algorithm performance. Moreover, our findings demonstrate that the proposed RSMA uplink strategy offers more flexible power allocation and significant performance improvement compared to conventional methods. 
{Future works can integrate channel prediction and MEC server selection modules into the multi-agent DRL framework and validate its real-time optimization capabilities under dynamic IRS configurations, scenarios with movable GUs, and systems with multiple MEC servers.} {In order to reduce the action space of decoding strategies, pointer networks and ordering learning frameworks for decoding sequences are worth studying in future studies.} {Further, to enhance the robust of the proposed algorithm, scenarios based on imperfect CSI will be prioritized in subsequent work.}

\bibliographystyle{IEEEtran}
\bibliography{shortref}

\end{document}